\documentstyle[preprint,epsfig,aps]{revtex}

\begin{document}

\author{Eduard Vives and Antoni Planes}

\address{Departament d'Estructura i Constituents de la Mat\`eria, \\
Facultat de F\'\i sica, \\
Universitat de Barcelona. \\
Diagonal, 647, E-08028 Barcelona, Catalonia}

\date{\today }

\title{Hysteresis and Avalanches in the Random Anisotropy Ising Model}

\maketitle

\tightenlines

\begin{abstract}

The  behaviour of  the Random  Anisotropy Ising  model at  $T=0$ under
local relaxation  dynamics is studied.  The model  includes a dominant
short-range   ferromagnetic  interaction   and  assumes   an  infinite
anisotropy at each site along local anisotropy axes which are randomly
aligned.  As a consequence,  some of the effective interactions become
antiferromagnetic-like and frustration  appears.  Two different random
distributions  of  anisotropy  axes   have  been  studied.   Both  are
characterized  by a  parameter that  allows control  of the  degree of
disorder in the system.  By using numerical simulations we analyze the
hysteresis   loop   properties   and  characterize   the   statistical
distribution of avalanches occuring during the metastable evolution of
the system  driven by an external field.   A disorder-induced critical
point is found in which  the hysteresis loop changes from displaying a
typical ferromagnetic  magnetization jump (large  avalanche spanning a
macroscopic fraction of the system) to a rather smooth loop exhibiting
only tiny avalanches.  The critical point is characterized by a set of
critical  exponents, which  are consistent  with the  universal values
proposed from the study of other simpler models.

\end{abstract}

\pacs{PACS numbers: 75.60.Ej, 75.30.Gw, 45.70.Ht, 05.50.+q}

\section{INTRODUCTION}
\label{section1}
Hysteresis occurs  in field-driven systems  showing history dependence
of   the  corresponding   response  \cite{Bertotti98}.    Disorder  is
acknowledged  to be  a  crucial ingredient  in determining  hysteresis
properties,  especially in the  so-called rate-independent  limit.  In
this case,  system properties do no exhibit  explicit time dependence.
For  this to  occur, two  conditions  must be  satisfied; (i)  thermal
fluctuations need to be irrelevant (athermal character) and (ii) field
driving rates must be small  enough.  Ideally, this corresponds to the
zero-temperature   quasi-static   limit.    Within   this   framework,
disordered systems  can be viewed  as described by  a multidimensional
energy landscape containing many  local metastable states separated by
large  energy barriers.   If thermal  fluctuations are  not operative,
these energy barriers  can only be overcome by  modifying the external
field  which tilts the  energy landscape.   The system  evolution thus
proceeds through jumps from one metastable state to another metastable
state  and therefore,  the  field-response shows  a discontinuous  and
apparent  stochastic  character.   In  magnetic  systems,  this  jerky
magnetization  response   corresponds  to  the   so-called  Barkhausen
noise\cite{Vergne81}. Moreover, similar behaviour has been reported in
many different systems,  including martensitic systems \cite{Vives94},
superconducting films  \cite{Wu95} and capillary  condensation systems
\cite{Lilly96} among  others.  All these systems exhibit,  as a common
feature,  a field-driven  first-order phase  transition  influenced by
disorder.   Actually, hysteresis  has its  origin in  this first-order
transition and  the characteristics of the hysteresis  loops depend on
both  the kind  and amount  of disorder  \cite{Dahmen96}.   With these
ideas in mind, different versions of lattice spin models with disorder
have  been  proposed  as  simple models  incorporating  the  essential
physics of these systems.  This includes, among other more complicated
models,  the  Random Field  (RFIM)  \cite{Sethna93},  the Random  Bond
(RBIM)     \cite{Vives94b}     and     the     Site-Diluted     (SDIM)
\cite{Tadic96,Obrado99} Ising  models.  Such models,  with appropriate
dynamics, show  rate-independent hysteresis and  associated Barkhausen
noise when  sweeping the  field at very  low temperature.  It  is also
found that the  system becomes magnetically softer when  the amount of
disorder  is increased.   This is  characterized  by a  change of  the
hysteresis  loops  from  sharp  to  smooth.  At  the  same  time,  the
distribution  of sizes  and durations  of Barkhausen  signals  is also
modified \cite{Vives94b}.  The striking  feature is the existence of a
specific  amount  of  disorder  at which  these  distributions  become
critical (power  law behaviour).  Therefore, the change  from sharp to
smooth hysteresis loops can be interpreted as a disorder-induced phase
transition, as  has recently been found experimentally  from the study
of Co/CoO films \cite{Berger00}.  The possible relationship of this
phase transition  (involving metastable states)  with a change  of the
ground state (equilibrium) of the system is still an open question.  A
number   of  recent   results  point,   however,  in   this  direction
\cite{Obrado99,Frontera99}.

In magnetic materials, a physically relevant source of disorder arises
from the  randomness of the magnetocrystalline  anisotropy. This paper
is aimed  at analysing  the intrinsic zero-temperature  hysteresis and
avalanche  properties in  systems with  such disorder.   Actually, the
random  uniaxial  single-site  anisotropy  Heisenberg model  has  been
considered to be  a suitable model to describe  magnetic properties of
amorphous alloys \cite{Fischer91}.  At mesoscales, this model can also
be  used,   as  a  reasonable  approximation,   to  describe  granular
materials, alloys  with precipitates, polycrystalline  systems etc...,
where   grains  form   single   magnetic  domains,   and  changes   of
magnetization  can   take  place  only   by  rotation  of   the  local
magnetization vector.   The model was first introduced  and studied in
the  mean  field  approximation  by Harris  et  al.   \cite{Harris73}.
Further   investigations   suggested   that   in  3$d$,   the   stable
low-temperature  phase is a  spin glass  phase, even  in the  limit of
strong anisotropy.  In this limit,  the spins are forced to be aligned
along the anisotropy axis and therefore, the model reduces to a Random
Anisotropy  Ising model  (RAIM) \cite{Derrida80}.   This is  the model
that will be considered in the present study.

To our knowledge,  the RAIM has not been studied  very much within the
above metastable evolution framework,  although it is a good candidate
to  display  a   disorder  induced  phase  transition  \cite{Vives99}.
Moreover,  it has  been argued  from symmetry  grounds that  it should
belong to the same universality  class as the RFIM \cite {Perkovic99}.
An interesting feature of this  model regards the fact that comparison
with experiments  may be considered  as being more realistic  than for
other more  idealized models such as  the RFIM or  the RBIM.  However,
modification, in a  controlled way, of the amount  of disorder in real
systems is not always easy to do.  For polycrystalline alloys changing
the  distribution of  anisotropy axes  is, in  principle,  possible by
impressing an orientation texture by application of a severe cold work
process  or   by  means   of  a  heat   treatment  (recrystallization)
\cite{Taylor61}.  In  the case of  magnetostrictive amorphous ribbons,
anisotropy   can  be   controlled  by   an  external   applied  stress
\cite{Livingston85,Appino99,Durin99}.   The  effect  of such  external
stress  is  to   decrease  the  amount  of  disorder   by  inducing  a
longitudinal anisotropy.

The paper is  organized as follows.  In the  next section we introduce
the model and  the details of the numerical  simulation procedure.  In
section  \ref{section3}  we present  the  results,  which include  the
description of  the hysteresis  loops, the Barkhausen  avalanches, and
critical phenomena.  Finally in  section \ref{section4} we discuss the
results and draw conclusions.

\section{Modelling}
\label{section2}

\subsection{Hamiltonian}
We  consider a  3d cubic  lattice with  spins ${\vec  S}_i$  and local
anisotropy  axes ${\hat  n}_i$ defined  at  each site  ($i=1, ...,  N=
L^3)$.  Each  unitary vector ${\hat n}_i$  is determined by  a pair of
spherical  polar coordinates  $(\theta_i, \phi_i)$.   Note  that since
such  unitary vectors  define a  direction  in space,  the angles  are
limited to the  ranges $0 < \phi_i <  2 \pi$ and $0< \theta_i  < \pi /
2$.   The axis  directions  are random  and independently  distributed
according  to a  generic density  probability  distribution $f(\theta,
\phi)$ which is normalized, i.e.:
\begin{equation}
\int_{0}^{2  \pi} d \phi  \int_0^{\frac{\pi}{2}} f(\theta,  \phi) \sin
\theta d \theta = 1.
\end{equation}
The general Hamiltonian of the system can be written as:
\begin{equation}
\label{general}
{\cal  H} = -\sum_{\langle  ij \rangle  }^{\frac{z}{2}N} J  {\vec S}_i
{\vec S}_j - \sum_{i=1}^N {\vec H} {\vec S}_i - D \sum_{i=1}^N \left (
{\hat n}_i {\vec S}_i \right ) ^2 - K \sum_{ij}^{N(N-1)} 3 \frac{(\vec
S_j \vec  r_{ij})(\vec S_i \vec  r_{ij})}{r_{ij}^5} - \frac  {\vec S_i
\vec S_j }{r_{ij}^3},
\end{equation}
\noindent  where  the  first  term  with  $J >  0$  accounts  for  the
ferromagnetic exchange energy, which is  short range and is assumed to
extend  only  to nearest-neigbour  pairs  ($z=6$  is the  coordination
number of the  cubic lattice). The second term  (Zeeman energy) stands
for the interaction with an external applied field $\vec H$. The third
term is the anisotropy energy.   Finally, the last term corresponds to
the long-range dipolar interactions that extend to all pairs $i, j$ of
the lattice.  The vector $\vec r_{ij}$ is the vector joining sites $i$
and $j$.

In the infinite anisotropy limit ($D >> 1$), the spins $\vec{S}_i$ are
constrained to lie on the $\hat n_i$ directions and therefore satisfy:
\begin{equation}
\vec{S}_i= {\hat n}_i S_i,
\end{equation}
where  $S_i$ is  now an  Ising spin  variable taking  values  $\pm 1$.
Under such conditions the Hamiltonian of the system can be written as:
\begin{equation}
{\cal H}=  - J \sum_{\langle  ij \rangle }^{\frac{z}{2}N}  S_iS_j \hat
n_i \hat n_j -\sum_i^{N} S_i \hat n_i \vec H - D N -K \sum_{ij}^{N(N-1)}  S_i S_j \frac{  3 (\hat n_j \hat r_{ij})(\hat  n_i \hat
r_{ij}) - (\hat n_i \hat n_j) }{r_{ij}^3}.
\label{hamiltonian}
\end{equation}
The dipolar  term represents, in  lattice models, a convenient  way to
describe the magnetostatic energy of  the system. Note that it changes
from being  pure ferromagnetic  for a pair  of interacting  spins with
${\hat n}_i  = {\hat n}_j  = {\hat r}_{ij}$ to  pure antiferromagnetic
when  ${\hat  n}_i =  {\hat  n}_j  \perp  {\hat r}_{ij}$.   The  exact
treatment of this term in numerical simulations is difficult since the
introduction  of  a  cut-off  in  the  interaction  range  can  induce
non-physical effects \cite{Kretschmer79}.  This term has been recently
considered within  the context of  athermal hysteresis studies  in the
case of perfectly  aligned anisotropy axes \cite{Szabo98,Magni99}. Its
main  effect on  the  hysteresis loop  is  to produce  a rather  large
nucleation jump  at the beginning of the  demagnetization process.  As
regards   its   influence  on   the   properties   of  the   avalanche
distributions, it has been argued \cite{Kuntz99} that its effect is to
cause mean-field behaviour.   Most of the results in  the present work
correspond  to  a  situation   with  rather  large  randomness.   Such
randomness and independence of  the orientation of the anisotropy axes
at  two  different  sites  $i$  and  $j$ is  expected  to  weaken  the
importance of the dipolar term.   Therefore, we have not considered it
in the present work.  The Hamiltonian with $K=0$ reads:
\begin{equation}
{\cal H}= - \sum_{\langle ij \rangle }^{\frac{z}{2}N}J_{ij} S_i S_j -H
\sum_i^{N} g_i S_i,
\label{hamK0}
\end{equation}
where the  first term is a random  bond term with $J_{ij}=  J \hat n_i
\hat n_j$  and the  second term describes  the random coupling  to the
external magnetic field (random  g-factor) with $g_i = \cos \theta_i$.
The constant term  $D N$ in Eq. (\ref{hamiltonian})  has been omitted,
since  it  represents  only a  shift  in  the  energy of  the  system.
Moreover, without loss of generality we will take $J=1$ as the unit of
energy.

In the present work we will  consider that the external field $\vec H$
keeps  its  direction  fixed along  the  $z$  axis  so that  only  its
magnitude  $H$ can  change.   Thus  the $g_i$  are  constant once  the
initial distribution of anisotropy axis has been quenched (notice that
rotation of $\vec H$ with fixed modulus could also be considered).

Finally,  it  should   be  remarked  that  the  second   term  in  Eq.
(\ref{hamK0}) acts,  for each  value of $H$,  as a random  field term.
This equivalence, nevertheless,  only applies for equilibrium studies.
The metastable evolution (hysteresis loops and sequence of avalanches)
of such  a random g-factor model  cannot, a priori, be  expected to be
equivalent to that of a RFIM.

\subsection{Modelling of disorder}
\label{modeling}
As   mentioned  in  the   introduction,  in   real  systems   such  as
polycrystalline alloys,  the distribution  of anisotropy axis  is very
much  dependent on  details of  the sample  preparation.  Here  we are
interested  in controlling  the amount  of disorder  in the  system by
controlling  the  distribution of  angles  $f(\theta,\phi)$.  We  have
considered two simplified models, which represent perturbations of the
two extreme cases of a  fully aligned anisotropy axis and a completely
random distribution;

\begin{itemize}

\item Model A

Uniform density within a cone $0<\theta<\theta_0$:
\begin{equation}
f(\theta, \phi)= \frac{1}{2 \pi \left ( 1- \cos{\theta_0} \right )} \;
h \left [ \theta_0 - \theta \right ],
\end{equation}
where  $h [x]$   is  the  Heaviside  step  function.    Note  that  for
$\theta_0=0$ the  axes are fully  aligned in the $z$  direction, while
for  $\theta_0=\pi/2$  the axes  are  isotropically distributed.   The
amount of disorder in the system increases with increasing $\theta_0$.

\item Model B

First-order correction to the isotropic distribution:
\begin{equation}
f(\theta,  \phi)= (1-\epsilon) \frac{1}{2  \pi} +  \epsilon \frac{3}{2
\pi} \cos^2 \theta.
\end{equation}
In this case the  parameter $\epsilon$ ranges within $-\frac{1}{2} \le
\epsilon  \le 1$  to ensure  that $f(\theta,  \phi)$ is  positive. For
$\epsilon > 0$ the distribution displays a peak at $\theta = 0$, which
flattens as $\epsilon$ goes to zero. Thus, in this case, the amount of
disorder increases with  decreasing $\epsilon$.  For $\epsilon=0$ this
distribution is uniform and  reduces to model A with $\theta_0=\pi/2$.
When  $\epsilon < 0$  the distribution  shows a  maximum at  $\theta =
\pi/2$, which corresponds to a tendency for the anisotropy axes to lie
isotropically  on a flat  disk perpendicular  to the  external applied
field.   Thus,  strictly speaking  when  $\epsilon$  decreases in  the
$\epsilon<0$   region   the   anisotropy   axes  orders   again,   but
perpendicular to  the external field. This  increases both frustration
and competing interactions.

\end{itemize}

\begin{figure}
\epsfig{file =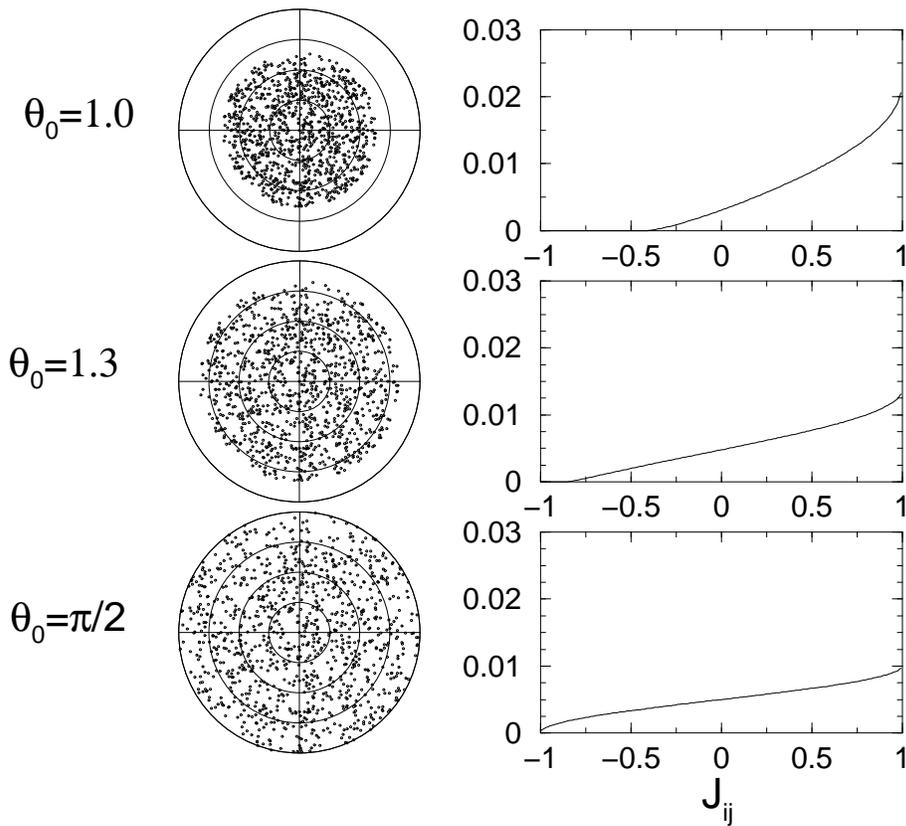, width=12cm}
\vspace{2cm}
\caption{Examples of the anisotropy  axis distributions of model A for
different  values  of  the  parameter $\theta_0$.   The  first  column
corresponds to the polar representation  described in the text and the
second  column to  the  distributions of  random  bonds (in  arbitrary
units).}
\label{FIG1}
\end{figure}

\begin{figure}
\epsfig{file=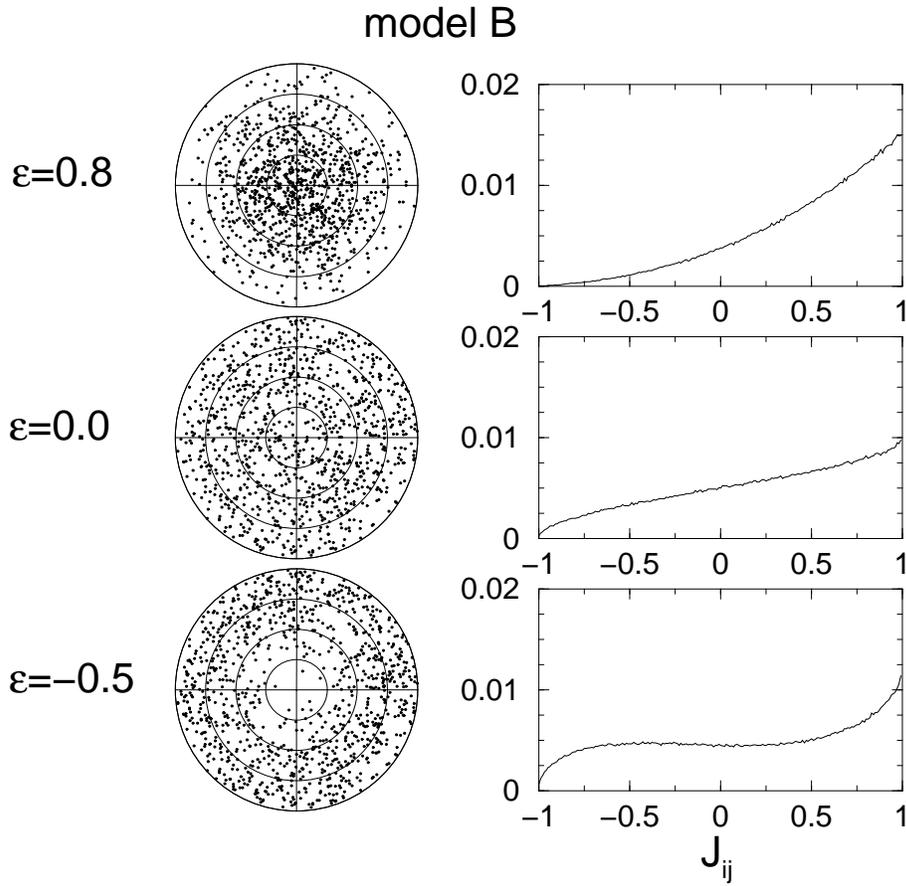, width=12cm}
\vspace{2cm}
\caption{Examples of the anisotropy  axis distributions of model B for
different  values  of  the  parameter $\epsilon$.   The  first  column
corresponds to the polar representation described in the text, and the
second column  to the corresponding distributions of  random bonds (in
arbitrary units).}
\label{FIG2}
\end{figure}

Figures \ref{FIG1}  and \ref{FIG2}  show a number  of examples  of the
anisotropy  axis   distribution  corresponding  to  models   A  and  B
respectively.  The  polar plot  in the first  column corresponds  to a
particular  sample  of  1000  anisotropy  axes  numerically  generated
according  to  $f(\theta,\phi)$.   The  polar  angle  in  these  plots
represents  $\phi$ while  the radius  represents $\sin  \theta$.  Note
that, with  this representation, a uniform  distribution of anisotropy
axes corresponds to a uniform  distribution of points on the circle of
unit radius.  For each  example we show the corresponding distribution
of exchange  interactions $J_{ij}= \hat  n_i \hat n_j $.   Moreover if
the   anisotropy   axes   are   restricted  to   $\theta<\pi/4$,   the
corresponding exchange interactions  are all ferromagnetic $J_{ij}>0$.
In contrast,  if $\theta>\pi/4$  in the case  of some  anisotropy axes
then the systems contains a fraction of antiferromagnetic bonds.

\subsection{Dynamics}

Regarding the  dynamics of the  system, two different  situations have
been discussed in  the literature \cite{Kuntz99,Zapperi98}.  The first
case  corresponds to  the  dynamics  of an  interface  or domain  wall
propagating   in   a   disordered   media   \cite{Zapperi98,Cizeau97}.
Actually, domain wall motion  is the predominant magnetization process
in the  approximately constant  permeability region of  the hysteresis
loops  \cite{Durin97}, and  therefore  such dynamics  is adequate  for
studies of the  linear region of the magnetization  curve.  The second
approach is  the so-called field-driven nucleation  proposed by Sethna
and co-workers  \cite{Sethna93}.  In  this case nucleation  and growth
are  treated  simultaneously. While  in  the  former  case only  spins
located at the propagating interface can flip, in the latter, any spin
of  the  system  can  turn  when such  a  flip  becomes  energetically
favourable. This  second approach seems more convenient  for the study
of the  full hysteresis  loops, in particular  in systems  with strong
local anisotropy and a lack of well-defined domain structure

The details of the dynamics used for the numerical simulations are the
following;  under slow  changes of  the external  magnetic  field, the
system  follows deterministic dynamics  corresponding to  local energy
relaxation. Due to  this local character, the evolution  of the system
will,  in general,  not follow  an equilibrium  path, but  rather will
evolve through metastable  states.  Actually, different configurations
of spins may correspond to the  same value of the external field. Such
different configurations are found depending on history conditions.

When  studying  the  full  hysteresis  loop,  the  starting  point  is
$H=\infty$  (or $H=-\infty$)  which corresponds  to the  unique stable
configuration  with   all  $S_i=1$  (or  $S_i=-1$).    We  proceed  by
decreasing (or increasing) the  field and compute, from (\ref{hamK0}),
the  change  of  energy  $(\Delta  {\cal H})_i$  associated  with  the
independent reversal of any spin $S_i$.  This change can be written as
\begin{equation}
(\Delta {\cal H})_i = 2 F_i S_i,
\end{equation}
where the local field $F_i$ acting on lattice site $i$ is given by;
\begin{equation}
F_i =   \sum_{n.n.} J_{ij} S_j + H \cos \, \theta_i. 
\end{equation}
\noindent The metastable states  correspond to those configurations of
spins for  which $\Delta {\cal H}_i >0  \; \; \forall i$.   When for a
certain value of $H$ one  of the spins become unstable ($(\Delta {\cal
H})_i=0$),  we  keep  $H$  constant  and flip  that  spin.   This  can
unstabilize some  neighbouring spins (for which  $(\Delta {\cal H})_i<
0$) which will be simultaneously flipped (synchronous dynamics).  This
is the origin of an avalanche.   Due to the fact that $\cos \theta_i >
0$,  when  decreasing (increasing)  the  field,  the  first spin  that
triggers  an  avalanche can  never  yield  an  increase (decrease)  of
magnetization.   However, such  inverse  magnetization reversals,  may
occur during the avalanche.  The procedure continues with $H$ constant
until  all the  spins become  stable  again. This  is the  end of  the
avalanche. The  external field is  then decreased (increased)  until a
new spin becomes unstable. Notice that the fact that the field remains
constant   during  the   avalanche  is   the  crucial   condition  for
rate-independent  hysteresis \cite{Bertotti98}.   It  is worth  noting
that  in our numerical  simulations we  have not  observed neverending
avalanches which  may, in general,  occur when using this  dynamics in
systems with  antiferromagnetic bonds.   Although we cannot  provide a
rigorous proof  for their absence,  we suspect that  such pathological
situations only occur for very special values of the random fields and
bonds  which have vanishingly  small probability  when the  angles are
distributed  continuously.   The  hysteresis  loops  are  obtained  by
measuring, as  a function of $H$,  the total magnetization  $M$ in the
$z$-direction defined as;
\begin{equation}
M = \sum_{i=1}^N S_i \cos{\theta_i}.
\end{equation}
Avalanches are characterized by  their duration and size. The duration
$t$ of the  avalanche corresponds to the number  of avalanche steps in
the algorithm  described above.  The avalanche size  can be quantified
in two different ways:

\begin{itemize}

\item The total change of  magnetization $\Delta M$ between the origin
and the end of the avalanche;
\begin{equation}
\Delta M  = \left  .  \sum  S_i \cos \theta_i  \right |_{end}  - \left
. \sum S_i \cos \theta_i \right |_{origin}.
\end{equation}
Note  that the  size of  the avalanches,  measured in  such a  way, is
bounded  by   twice  the   saturation  magnetization  of   the  system
$M_{sat}=\sum_{i=1}^N \cos \theta_i$.

\item The total number of  spins flipped during the avalanche. We will
denote such a magnitude by $s$.  In this case, as opposed to the above
definitions, the avalanche size $s$  is not bounded by the system size
$L^3$ due to the possibility of inverse spin flips.

\end{itemize}

\section{Results}
\label{section3}

In  this  section  we  present  the  main  results  of  the  numerical
simulations.   We  have  studied  3d systems  with  periodic  boundary
conditions and sizes $L=6,10,20$,$30$ and $40$.

\subsection{Hysteresis loops}

\begin{figure}
\epsfig{file=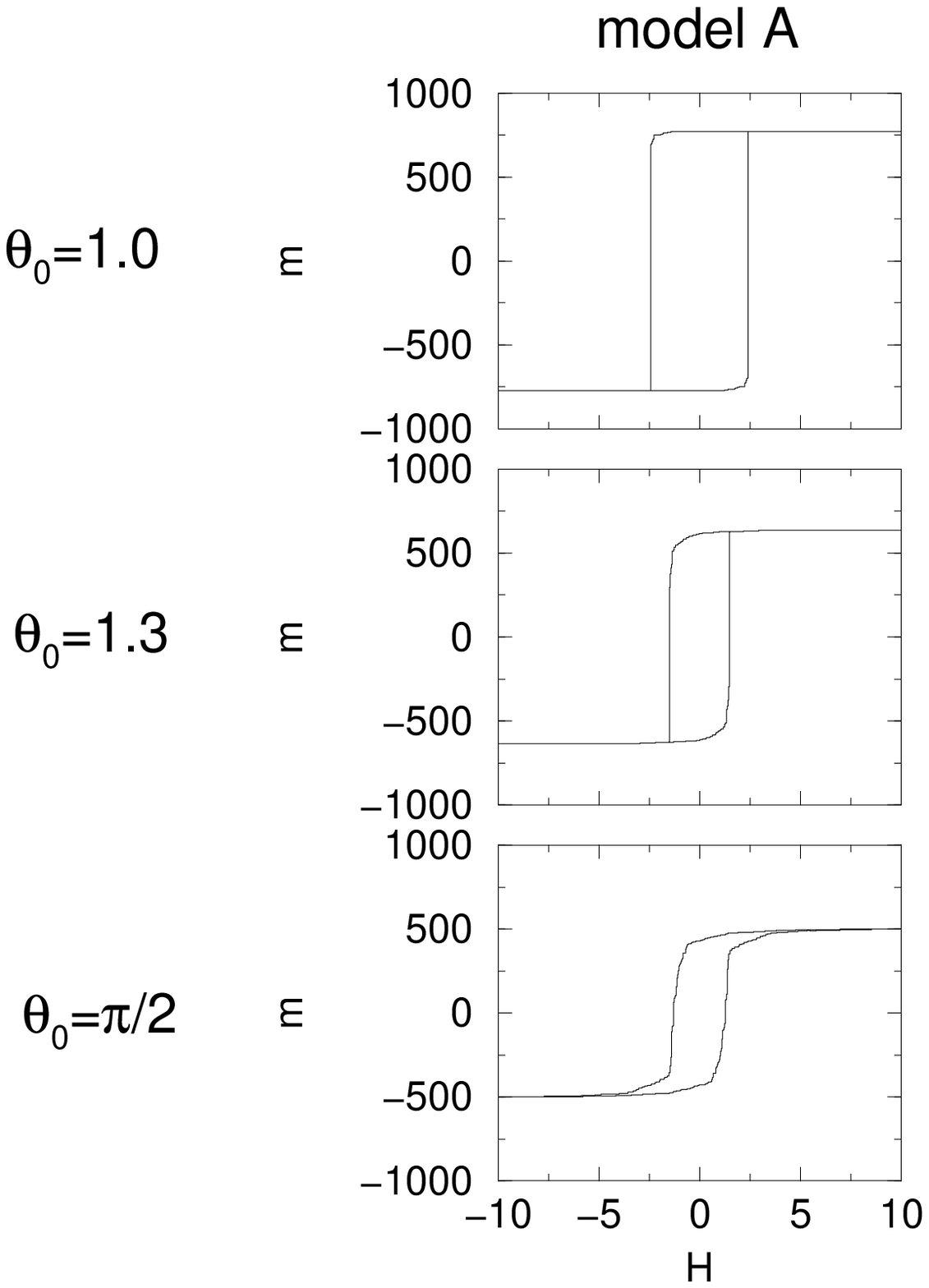, width=12cm}
\caption{Examples of hysteresis loops  of model A for different values
of  the   parameter  $\theta_0$.   Data  correspond   to  a  numerical
simulation of a system with $L=10$.}
\label{FIG3}
\end{figure}

\begin{figure}
\epsfig{file=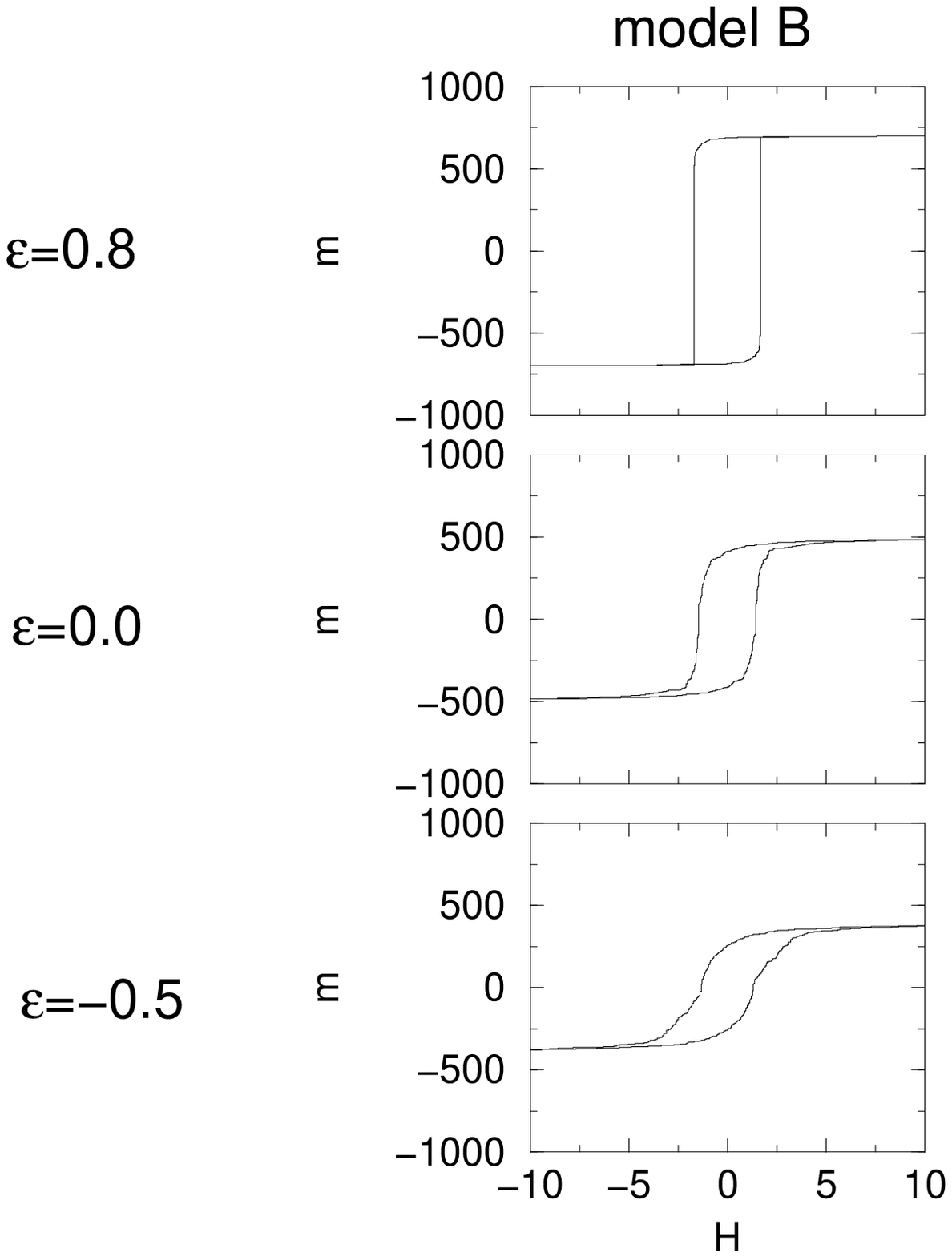, width=12cm}
\caption{Examples of hysteresis loops  of model B for different values
of  the   parameter  $\epsilon$.   Data  correspond   to  a  numerical
simulation of a system with $L=10$.}
\label{FIG4}
\end{figure}

Figures  \ref{FIG3} and  \ref{FIG4}  show examples  of the  hysteresis
loops  corresponding to  models A  and  B (system  size $L=10$),  with
different amounts of disorder (controlled by the parameters $\theta_0$
and $\epsilon$  as indicated).  First  of all, it should  be mentioned
that the loops are symmetrical  with respect to changes $H \rightarrow
-H$ and  $ M\rightarrow  -M$.  This property  comes directly  from the
symmetry of the Hamiltonian (\ref{hamK0}).

For  both models the  qualitative morphological  changes of  the loops
when disorder increases are very similar.  Figure \ref{FIG5} shows the
dependence  of the  saturation magnetization  $M_{sat}$,  the remanent
magnetization $M_{rem}$,  the coercive  field $H_{coe}$ and  the total
dissipation (area within the loop) $W$ on the parameter ($\theta_0$ or
$\epsilon$)  controlling  the  amount  of  disorder.   In  both  cases
$M_{sat}$,    $M_{rem}$,   and    $W$    decrease   with    increasing
disorder. Notice that $M_{sat}$ can be obtained exactly by integration
of $f(\theta,\phi)  cos(\theta)$ over all the  spatial directions. The
most remarkable  difference concerns  the behaviour of  $H_{coe}$.  In
model A it exhibits a monotonous decrease with increasing disorder. In
model  B  after the  initial  decrease  of  $H_{coe}$ with  increasing
disorder,  a minimun is  reached; for  negative values  of $\epsilon$,
$H_{coe}$  increases.  From  these  averaged morphological  magnitudes
apparently no signs of singular  behaviour as a function of the amount
of disorder  is found.  The  detection of a  possible disorder-induced
critical  point needs  a detailed  study of  avalanches which  will be
presented in the next subsection.

It  is illustrative  to show  a sequence  of snapshots  of  the system
configuration during  a demagnetization  process.  These are  shown in
Fig.   \ref{FIG6}  which  corresponds  to  model  A  with  $L=20$  and
$\theta_0=1.3$.  Black indicates the lattice sites with reversed spins
$S_i=-1$.   The simulation  starts from  saturation (all  the $S_i=1$)
with  a  very  large  applied  field.   The  different  configurations
correspond to the same plane  (of the 3d system) for decreasing values
of the external field $H=  -0.10, -1.10, -1.44, -1.48, -1.49$.  During
the  first stages  of  the demagnetizing  process  the main  dynamical
mechanism  is the  nucleation of  the reverse  magnetization  phase by
flipping isolated spins.  In contrast, in the middle of the hysteresis
loop the evolution takes place  by growth (depinning) of such domains.
This occurs  by means of  large avalanches which produce  reversals of
large fractions of the system.

Prior to the analysis of such avalanches it is interesting to consider
another  feature of  the hysteresis  loops.  The  analysis  of partial
cycles enables  study of the  existence of the so-called  return point
memory  (RPM)  property.   The  mathematical  conditions  for  such  a
property to  occur have been  discussed for the  RFIM \cite{Sethna93},
the  RBIM \cite{Vives94} and  the SDIM  \cite{Obrado99}. The  two RAIM
models (A and B) studied in  this work do not exhibit such a property,
except for  those situations  in which no  effective antiferromagnetic
interactions occur.   Figure \ref{FIG7}  shows an example  of internal
loops revealing the  failure of the RPM property.  This  is due to the
existence  of  reverse spin  flips  during  the  avalanches, and  even
reverse avalanches  that occur for  large enough amounts  of disorder.
It  is worth  mentioning that  such  reverse flips  represent a  small
fraction of  the total  number of flips.   For instance, close  to the
critical amount of disorder (defined below) it represents less than $5
\%$.

\subsection{Barkhausen avalanches}

The statistical  analysis of the avalanches is  performed by measuring
its  size $s$  and duration  $t$, in  a half  hysteresis  loop. Figure
\ref{FIG8} shows  the probability $p(s)$ of occurence  of an avalanche
of  size $s$  for model  $A$ and  increasing values  of the  amount of
disorder $\theta_0$.  Data, represented in log-log plots correspond to
the analysis of $300$ different half-loops with different realizations
of the disorder.  For small values of $\theta_0$, the cycles contain a
very large avalanche giving a peak  on the right-hand side of the plot
and  a  certain fraction  of  small  avalanches,  on the  left.   This
behaviour is  called ``supercritical''.  On the other  hand, for large
values of $\theta_0$ the system behaves ``subcritically'' showing only
small  avalanches.  For  an intermediate  value  $\theta_{0}^c(L=30) =
1.44  \pm 0.01$, the  distribution of  avalanches becomes  a power-law
(``critical'') characterized  by an exponent $\tau' =  2.06 \pm 0.05$.
The details of the exponent  fitting procedure together with the study
of the dependence with the finite size of the system will be presented
in  the next  section.   Figure \ref{FIG9}  shows  the avalanche  size
distribution $p(s)$ for model $B$.  In this case critical behaviour is
found  at  $\epsilon^c(L=30)  \sim  0.2$  with  a  power-law  exponent
$\tau'=2.10\pm 0.05$

It  is interesting to  consider the  following two  remarks concerning
such  avalanche size  distributions. On  the one  hand,  the avalanche
analysis could also be  performed by characterizing the avalanche size
by $\Delta M$, instead of  $s$.  Figure \ref{FIG10} shows a comparison
of the  two histograms (number  of avalanches versus size)  $N(s)$ and
$N(\Delta M)$  in the case of  model A with $L=20$,  $\theta_0 = 1.39$
and averages over $300$ different realizations.  The agreement between
both histograms is very good.  Even in the small avalanche size region
both  histograms exhibit  almost  the same  behaviour indicating  that
effects  related  to  the  existence  of  inverse  jumps  are  totally
negligible.   Therefore, from  now on,  we will  only  consider $p(s)$
distributions.   (Note  that   in  figure  \ref{FIG10}  the  histogram
$N(\Delta  M)$ has  been constructed  by taking  bins of  size $\Delta
M=8$.   This  is not  necessary  for $N(s)$  since  $s$  is a  discret
variable.  Thus,  for the sake of comparison,  the histogram $N(\Delta
M)$ must be displayed after being divided by a factor $8$).

On the  other hand, it should  be mentioned that  the large avalanches
occuring in the  supercritical case span the full  simulated system at
least in one direction.  This is illustrated in Fig. \ref{FIG11} which
corresponds to the avalanche distribution  of model A with size $L=20$
and  $\theta_0=1.39$.  For this  amount of  disorder the  system still
behaves slightly supercritically.   The three histograms correspond to
the  distribution of  all  the avalanches  (bottom), the  non-spanning
avalanches (middle) and the spanning  ones (top).  The inset shows the
histogram  of spanning  avalanches on  a linear  scale,  revealing the
existence  of a  characteristic  size, which  increases when  disorder
decreases and  system size increases.  Actually,  in the thermodynamic
limit  such   spanning  avalanches  will  be   infinite,  involving  a
macroscopic   fraction  of  the   the  system   and  giving   rise  to
magnetization  discontinuities in  the hysteresis  loop.  It  is worth
noting  that in  previous  studies of  the  same problem  in the  RFIM
\cite{Dahmen96}  such spanning  avalanches were  substracted  from the
histograms for the analysis of the critical behaviour.  In the present
work  we have  decided to  keep  them since,  as will  be seen,  their
occurence provides a criteria for locating the critical point.

\subsection{Criticality}

The power-law  behaviour of  the avalanche size  distributions reveals
the  existence of  criticality in  the system.   For this  reason such
transitions related to the change of properties of the hysteresis loop
and of  the Barkhausen noise  distribution when disorder  is increased
are  called   disorder-induced  critical  points.    They  share  many
similarities with the classical  critical points, but one should never
forget that a number of features are different; firstly we are dealing
with a history-dependent metastable  evolution of the system, i.e.  an
out-of-equilibrium   problem,  thus  many   thermodynamical  equations
relating     critical     exponents,      may     not     be     valid
\cite{Maritan93,Frontera00}.   At this  point it  should  be mentioned
that for the RFIM, for which the exact equilibrium trajectories can be
obtained, it has been found numerically that a transition point exists
for the same amounts of disorder in equilibrium.  \cite{Frontera00}. A
second  remark  concerns   the  fact  that  we  are   dealing  with  a
deterministic  phenomenon  at  $T=0$  and thus  fluctuations  (in  the
standard  sense), do not  exist.  By  studying systems  with different
realizations  of the  disorder corresponding  to the  same probability
distribution $f(\theta,\phi)$,  one can  define average values  of any
generic property $z$  that we will denote as  $\langle z \rangle$.  We
can also  define ``fluctuations'' as  $\langle z^2 \rangle-  \langle z
\rangle^2$,   but  the   extrapolation  of   these  averages   to  the
thermodynamic limit may hide some mathematical inconsistencies.

The  consequences arising  from the  two remarks  above are  still not
totally understood.  For  instance, for such disorder-induced critical
points it is not clear what  the order parameter is. One choice is the
system magnetization  per site $\langle  M/L^3 \rangle$. Nevertheless,
the   fact   that   the    system   displays   hysteresis   for   both
$\theta_0<\theta_0^c$ and  $\theta_0>\theta_0^c$ implies that $\langle
M/L^3 \rangle  $ does not go to  zero at the critical  point.  For the
RFIM, Sethna and coworkers  \cite{Dahmen96} use $\langle M/L^3 \rangle
-\langle  M_c/L^3   \rangle$  (where  $M_c$   is  the  value   of  the
magnetization  at the  critical point).   Besides the  fact  that this
quantity does  not remain equal to  zero above the  critical point, it
adds to the  problem of determining $M_c$. A  second choice, which was
originally  used for  the study  of  the RBIM  \cite{Vives94b}, is  to
measure the size  $\langle s_{max} \rangle $ of  the largest avalanche
in the hysteresis loop. Clearly this  is a quantitiy that for a finite
system is not a suitable order  parameter since it never goes to zero.
However, for  the infinite system, $\langle  s_{max}/L^3 \rangle$ will
be  zero for  any degree  of disorder  except for  those for  which an
avalanche spanning  a macroscopic portion  of the system  occurs. This
leads  to the  existence of  a discontinuity  in the  hysteresis loop.
Thus, in the present paper, we  have chosen this quantity as the order
parameter.

Figure  \ref{FIG12}  displays the  behaviour  of $\langle  s_{max}/L^3
\rangle  $ as  a function  of $\theta_0$  for different  system sizes.
Data  corresponds to  averages  over  $50, 200,  300,  500$ and  $300$
different  realizations of  the disorder  for $L=40,30,20,10$  and $6$
respectively.  The  (pseudo-) phase  transition for the  finite system
will correspond  to the  inflection point of  such curves.   The exact
location of $\theta_0^c(L)$ can be obtained, for instance, by means of
a 4rth-order  polynomial fitting of  the inflection point or,  after a
numerical derivative,  a 2nd-order polynomial fitting  of the maximum.
This  gives two  slightly different  estimations of  the  the critical
point.   An independent  way of  locating the  phase transition  is to
measure the  duration $t_{max}$ of  the longest avalanche in  the half
hysteresis loop.   The average of  such a quantity  $\langle t_{max}/L
\rangle$  is  also  shown  in  Fig.   \ref{FIG12}  as  a  function  of
$\theta_0$ for different system  sizes ($t_{max}$ is normalized by $L$
since this is the minimum number of steps needed in order to cover the
full  system).  This  quantity displays  a maximum  at $\theta_0^c(L)$
which  is  also  fitted  by  using  a  2nd-order  polynomial.   Figure
\ref{FIG12B} shows  equivalent data to that of  Figure \ref{FIG12} for
model B. In this case only systems up to $L=30$ have been studied. The
simulation of larger systems in  this case is much more time consuming
than for model A due to the wider distribution of anisotropy axes.

A fourth  method for the location  of the critical  point results from
the quantitative analysis of  the avalanche size distribution of Figs.
\ref{FIG8}  and  \ref{FIG9}.  These  distributions  correspond to  the
statistical analysis of all the avalanches occuring in full half-loops
for many realizations of disorder.  In general they are well fitted by
an   exponentially   corrected   power-law  probability   distribution
\cite{Vives99};
\begin{equation}
\label{powexp}
p(s; \tau', \lambda) = \frac{1}{A} s^{-\tau'} e^{-\lambda s},
\end{equation}
\noindent  where  $A$   is  not  an  extra  free   parameter  but  the
normalization  factor.  As  mentioned  before the  avalanche size  $s$
takes  discrete values  and, strictly  speaking, is  not  bounded from
above due to the possibility of inverse flips.  For the computation of
the  normalization  factor  $A$  we  have  chosen  the  largest  value
$s_{max}$ of each set of data, which in all cases has been found to be
lower  than $L^3$.   Thus  $A$ (which  is  a function  of $\tau'$  and
$\lambda$) is given by:
\begin{equation}
A(\tau',\lambda)= \sum_{s=1}^{s_{max}} s^{-\tau'} e^{-\lambda s}.
\end{equation}
\noindent  The fits are  performed by  the maximum  likelihood method,
which  is  independent  of  any  binning  process  or  representation.
Examples of the  fits are also shown in Fig.  \ref{FIG8}. As a general
comment, it  should be mentioned that  the fits are very  good for the
subcritical, critical  and slightly supercritical  distributions.  For
the deep supercritical distributions they are not that good due to two
different problems;  (i) the existence of large  avalanches which span
an important fraction of the  system makes it difficult to have enough
statistics   and  (ii)  the   fact  that   the  proposed   model  (Eq.
\ref{powexp}) is not well suited to describe the occurence of the peak
(with a certain characteristic size) in the large $s$ regions.

For model A the values obtained of $\lambda$ and $\tau'$ as a function
of  $\theta_0$ are  shown in  Fig.  \ref{FIG13}  for  different system
sizes ($L=10, 20$ and $30$).  For small amounts of disorder $\theta_0<
\theta_0^c$,  one gets $\lambda<0$.   For $\theta_0>  \theta_0^c$, one
gets $\lambda>0$.   The estimation of $\theta_0^c(L)$  can be obtained
by interpolating the value  of $\theta_0$ for which $\lambda(\theta_0)
= 0$. This, nevertheless,  shows large uncertainites that increase for
increasing values of $L$.  The  same kind of analysis can be performed
with the corresponding similar results  for model B. They are shown in
Fig.  \ref{FIG13B}.

The  four  estimations above  of  $\theta_0^c(L)$  are  shown in  Fig.
\ref{FIG14} as a  function of $L^{-1}$.  The results  exhibit a strong
dependence  on the  size $L$  of the  simulated system,  as  occurs in
numerical simulation  of standard critical  phenomena. As can  be seen
the  four estimations  of $\theta_0^c$  decrease with  increasing $L$.
Except for the data obtained  from the analysis of the distribution of
avalanches which  shows large error bars, the  linear extrapolation to
$L  \rightarrow  \infty$  indicates  a  compatible  common  value  for
$\theta_0^c(L\rightarrow \infty)$.

The exact treatment of the  dependence of the measured quantities with
$L$   must  be   performed   within  the   framework  of   finite-size
scaling. \cite{Kim93}.

\subsection{Finite-size scaling}

According  to  the  standard  finite-size  scaling  hypothesis,  as  a
function of system size $L$, $s_{max}$ and $t_{max}$ behave as:
\begin{equation}
s_{max}(x,L)/L^3   \sim   L^{-    \frac{\beta}{\nu}}   F_s   \left   (   x
L^{\frac{1}{\nu}} \right )
\end{equation}
\begin{equation}
t_{max}(x,L)  \sim L^{\frac{z}{\nu}} F_t  \left (  x L^{\frac{1}{\nu}}
\right ),
\end{equation}
\noindent where $x$ is the reduced amount of disorder. For the case of
model A,  $x=(\theta_0 - \theta_0^c(L))/  \theta_0^c(L)$. The functions
$F_s$ and $F_t$  are scaling functions and $\beta$,  $z$ and $\nu$ are
critical exponents.  The different estimations of the critical amounts
of disorder  $\theta_0^c(L)$ for the  finite system should  also scale
with $L$ as:
\begin{equation}
\theta_0^c(L) - \theta_0^c(\infty) \sim L ^{-\frac{1}{\nu}}.
\end{equation}
\noindent There are different ways  to fit the four exponents $\beta$,
$\tau'$, $z$ and $\nu$  and determine $\theta_0^c(\infty)$.  Since the
behaviour  of  $\theta_0^c(L) $  is  quite  linear  with $L$  in  Fig.
\ref{FIG14},  this  suggests  that  to  a first  approximation  it  is
reasonable to take $\nu \sim 1$.  This justifies the linear fits shown
in Fig. \ref{FIG14}. A value consistent with all the extrapolations is
$\theta_0^c(\infty) =  1.33 \pm  0.03$ (indicated by  an arrow  on the
vertical axis).   With this estimation of  $\theta_0^c(\infty)$ we can
refine the  value of $\nu$ by  performing a linear fit  to the log-log
plot  of  $  \left  (  \theta_0^c(L)- \theta_0^c(\infty)  \right  )  $
vs. $L$. The obtained value is $\nu = 1.0 \pm 0.1$

Once  $\theta_0^c(\infty)$  and $\nu$  are  determined, the  exponents
$\beta$ and  $z$ can be obtained  by analyzing the change  with $L$ of
the  height   and  slope  at   the  inflection  point  in   the  curve
$s_{max}(\theta_0,L)$ and  the height and curvature at  the maximum in
$t_{max}(\theta_0,L)$.   From   linear  fits  to   log-log  plots  the
following  estimations  are  obtained:  $\beta  =  0.06  \pm  0.05  $,
$-\beta+1/\nu = 0.8  \pm 0.1$, $z=1.6\pm0.02$ and $z+2/\nu=3.3\pm0.2$.
Such values are consistent with  a final estimation of $\beta= 0.1 \pm
0.1$ and $z=1.6 \pm 0.1$.  The  goodness of the final set of exponents
can  be finally  tested by  plotting  the scaling  fuctions $F_t$  and
$F_s$, which  are shown  in Fig.  \ref{FIG15}. Within a  rather good
approximation data  collapses onto a single  curve, which demonstrates
the assumed scaling  hypothesis.  

A  similar analysis has  been performed  for model  B.  In  this case,
instead of fitting  a new set of exponents we have  tried to scale the
data in Fig. \ref{FIG12B} with the set of exponents obtained above for
model  A.    The  resulting  scaling  functions  are   shown  in  Fig.
\ref{FIG15B}. Again a good data collapse is obtained, demostrating the
validity of  the scaling hypothesis for  model B with the  same set of
critical exponents.

A   summary   of   the    exponents   found   are   given   in   Table
\ref{TABLE1}. Values  corresponding to other 3d  models and mean-field
calculations are  also presented for  comparison.  This point  will be
discussed in section \ref{section4}.

\subsection{Critical field}

The avalanche  size distributions  analyzed in the  previous sections,
corresponds to the study of the whole hysteresis cycle.  Nevertheless,
the  simulations of  the  RFIM \cite{Sethna93}  suggested  that it  is
convenient to  analyze such distributions  at different points  of the
hysteresis loop.  Strictly, criticality is expected to occur only at a
certain value of  the field $H^c$ (critical field).   In this case the
power-law  distribution  of avalanche  sizes  is  characterized by  an
exponent $\tau$ which for the RFIM takes a value $\tau = 1.6\pm 0.06 $
and  is  related  to   $\tau'$  through  a  certain  scaling  relation
\cite{Perkovic99}.

The  study of  $p(s)$  at $H_c$  is  quite difficult  since to  obtain
sufficiently accurate statistics  for a given value of  $H$ requires a
large number of realizations  of disorder.  Fig.  \ref{FIG16} presents
such an  analysis for model $A$ in  the case of a  system with $L=20$,
$\theta_0=1.39$  ($\sim  \theta_0^c(L=20)$)  and averages  over  $300$
realizations.  The  distributions have been computed  by analysing the
avalanches occuring  in windows  of size $\Delta  H = 0.5$  around the
indicated   values  of   the   external  applied   field  during   the
demagnetizing process.   The critical distribution occurs  for a field
at $|H^c(L=20)|  \sim -1.5$.  For  values of $H$  significantly larger
and  smaller,  the   distributions  exhibits  an  evident  exponential
damping.   In principle,  a more  quantitative treatment  is possible,
which consists of fitting the  data with the distribution given by Eq.
\ref{powexp} (replacing  $\tau'$ by  $\tau$).  Results for  $\tau$ and
$\lambda$  are shown  in Fig.\ref{FIG17}  as a  function of  $H$.  The
figure reveals the  existence of a critical region  with $\lambda \sim
0$ and $\tau \sim 1.5$.  It is worth noting that outside this critical
region, the  fit of  Eq.  \ref{powexp} renders  values of  $\tau$ well
below the  critical value.  This  method of determining $H^c$  is very
approximate,  since the need  for large  enough statistics  requires a
large field window that introduces considerable bias.

Finally, it is interesting to  compare $H_{coe}$ with the value of the
field $H_{s_{max}}$  at which the largest avalanche  ($s_{max}$) for a
demagnetizing  process (from  positive  $H$ to  negative $H$)  occurs.
Experimentally, in  the region of large disorder  $H_{s_{max}}$ can be
determined by locating the  field for which the macroscopic hysteresis
loop  exhibits maximum  slope, i.e.   maximum  susceptibility.  Figure
\ref{FIG18}  compares $H_{s_{max}}$  and  $H_{coe}$ as  a function  of
$\theta_0$  for three different  system sizes  $L=10$, $20$  and $30$.
Data corresponds  to averages over $1000, 300$  and $300$ realizations
of  disorder respectively.  As  expected, for  low values  of disorder
both  $H_{s_{max}}$  and  $H_{coe}$  coincide; the  largest  avalanche
associated with the magnetization reversal crosses the line $M=0$ and,
thus,  determines  $H_{coe}$.   In  contrast,  for  large  amounts  of
disorder the  largest avalanche  in the hysteresis  loop occurs  for a
value of the field more  negative than $H_{coe}$.  The inset in figure
\ref{FIG18}  shows  the   actual  distribution  of  $H_{s_{max}}$  and
$H_{coe}$  over different  realizations of  disorder and  $L=10$. Note
that both distributions are quite Gaussian and that for large disorder
are split  in such  a way that  the distribution of  $H_{coe}$ remains
rather  sharp  while   the  distribution  of  $H_{s_{max}}$  broadens.
Whether  or  not  the   coincidence  of  $H_{coe}$  and  $H_{s_{max}}$
determines  the   critical  field  is   a  question  that   cannot  be
definitively answered from our results.

\vspace{5mm}

\section{Discussion}
\label{section4}

In this  section we  compare our results  with those  corresponding to
other models and experiments reported in the litterature.

In the present RAIM, hysteresis arises from energy barriers separating
metastable states  which have their  origin in the  effective coupling
between spins.  This effective coupling  is modified by changes in the
distribution of anisotropy axes, but  even in the absence of disorder,
(corresponding   to  the   zero-temperature   standard  Ising   model)
hysteresis occurs. In magnetism,  hysteresis can be interpreted within
the  framework  of  the Stoner-Wohlfarth  model  (SWM)\cite{Stoner48}.
This model  gives an  essentially different description  of hysteresis
than that  proposed in  this paper.  For  the SWM,  independent single
magnetic  domains with  continuously orientable  magnetic  moments are
considered.  These single domains can  be identified with the spins in
the  present  model.   Hysteresis,  in  the SWM,  arises  from  energy
barriers  originating from the  competion between  uniaxial anisotropy
and Zeeman energy.  Actually, no  hysteresis occurs in the SW model in
the infinite anisotropy limit.

\begin{figure}
\epsfig{file=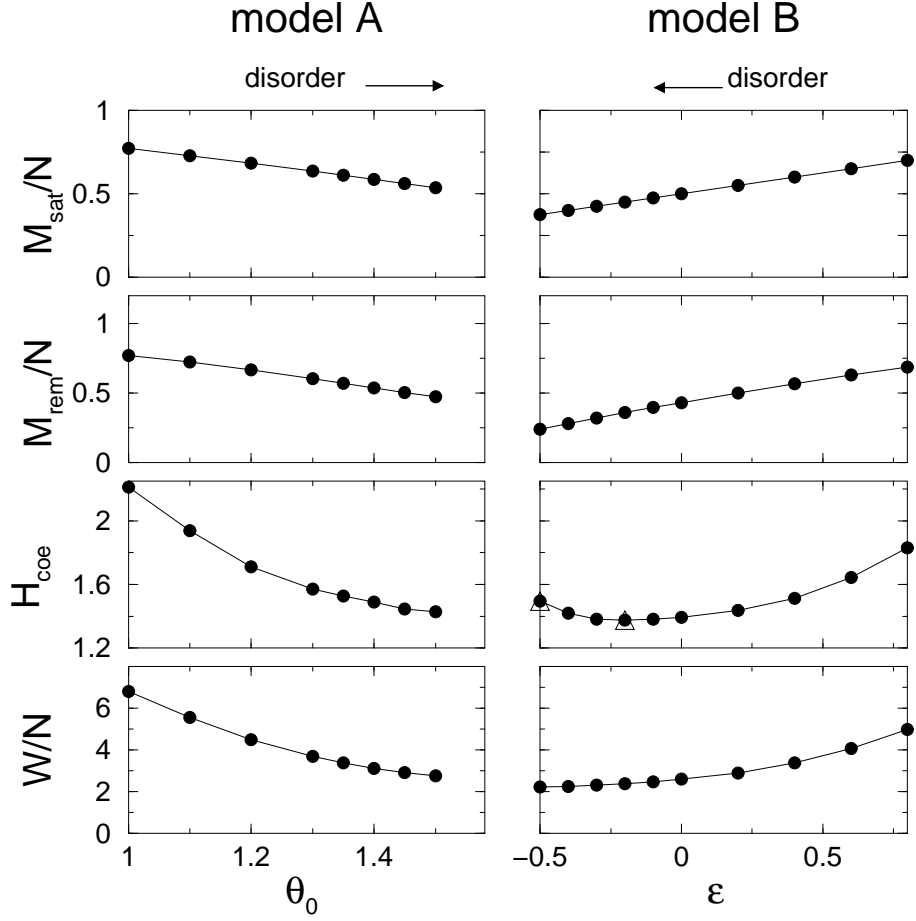, width=12cm}
\caption{Evolution of different morphological properties of hysteresis
loops for models A (first column)  and B (second column) as a function
of   disorder:    saturation   magnetization   ($M_{sat}$),   remanent
magnetization ($M_{rem}$), coercive  field ($H_{coe}$) and dissipation
($W$). Data  corresponds to  averages over 100  hysteresis loops  of a
system with $L=10$,  except for the open triangles  that correspond to
$L=20$.   In the  two  top figures,  corresponding  to $M_{sat}$,  the
continuous  lines  show  the  exact  analytical  calculations,  giving
$\sin^2(\theta_0)/(2  (1-\cos(\theta_0))$ and  $1/2 +  \epsilon/4$ for
models A and B respectively. In all cases, error bars are smaller than
symbol sizes.}
\label{FIG5}
\end{figure}

\begin{figure}
\epsfig{file=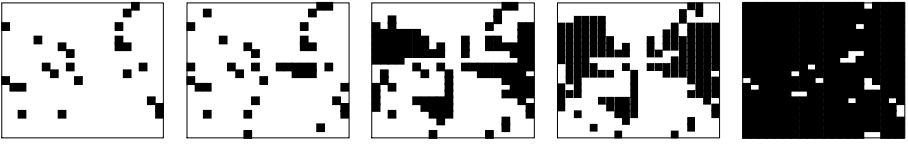, width=12cm}
\vspace{2cm}
\caption{Sequence   of   snapshots   of   the   system   configuration
corresponding to  a numerical  simulation of model  A with  $L=20$ and
$\theta_0=1.3$.  The  picture shows the same  section perpendicular to
the $[001]$ direction of the 3d system for different values of $H$}
\label{FIG6}
\end{figure}

\begin{figure}
\epsfig{file=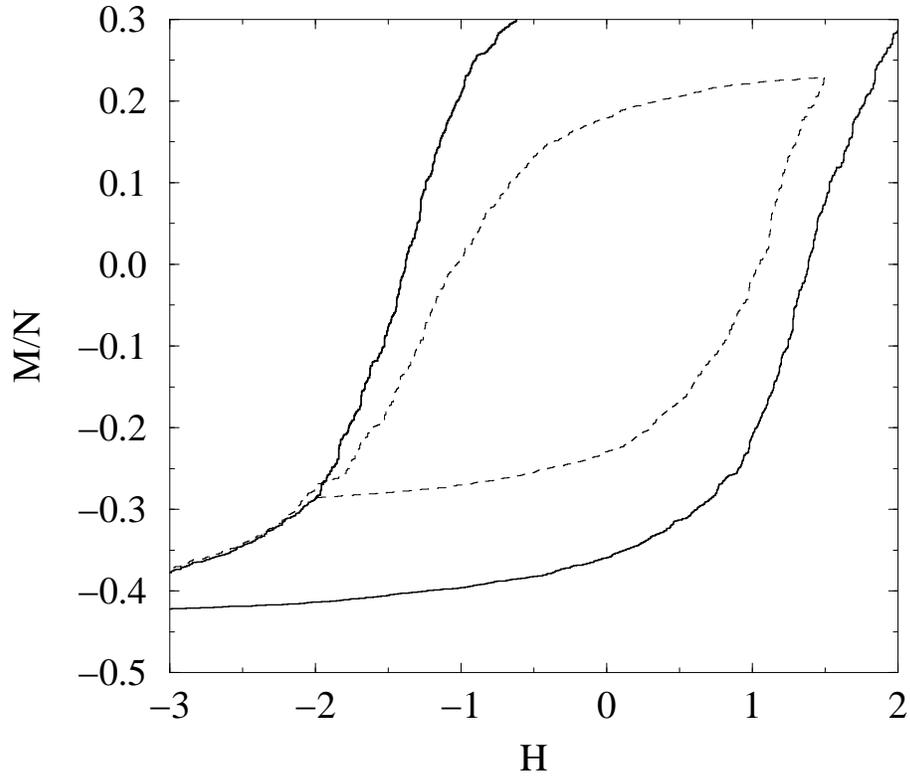, width=12cm}
\caption{Example of a  hysteresis loop of model B  showing the failure
of  the return  point  memory property  after  partial cycling.   Data
corresponds  to a  simulation  of a  system  with $\epsilon=-0.2$  and
$L=10$.}
\label{FIG7}
\end{figure}

\begin{figure}
\epsfig{file=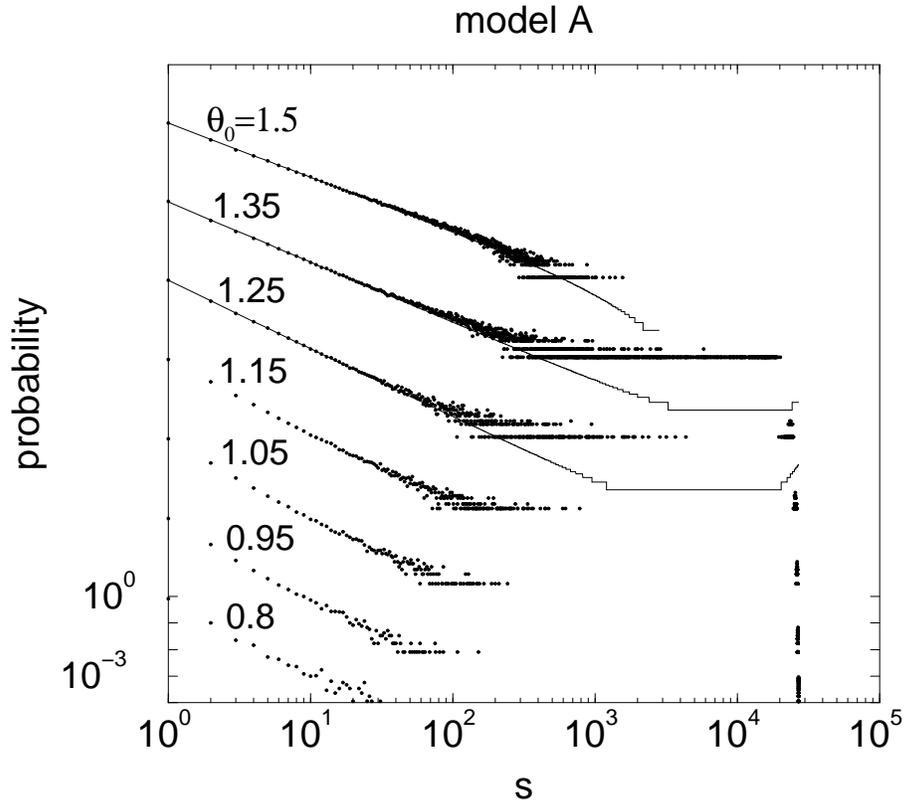, width=12cm}
\caption{Probability  distribution $p(s)$  of avalanche  sizes  in the
hysteresis loop  of model  A.  Data corresponds  to an average  of 300
runs for  a system with $L=30$  and different values  of $\theta_0$ as
indicated.   Except  for  the  bottom  curve,  the  curves  have  been
vertically  shifted  (three decades  each)  in  order  to clarify  the
picture.   Continuous lines  correspond  to examples  of  the fits  of
equation (\ref{powexp}).}
\label{FIG8}
\end{figure}

\begin{figure}
\epsfig{file=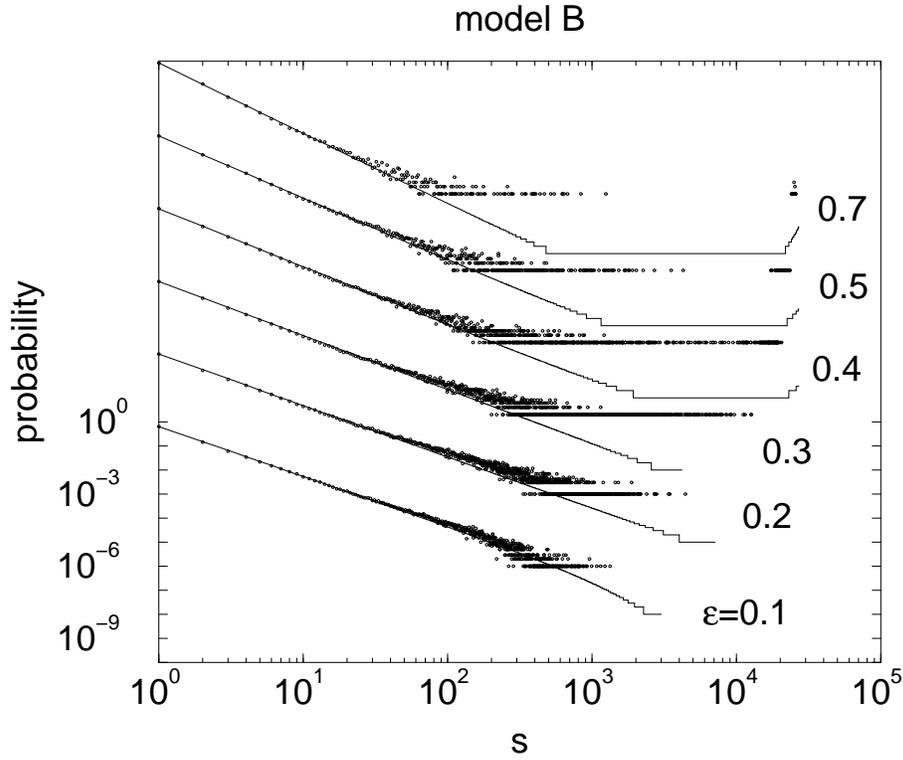, width=12cm}
\caption{Probability distribution of avalanche sizes in the hysteresis
loop of  model B.  Data  corresponds to an  average of 100 runs  for a
system with  $L=30$ and different  values of $\epsilon$  as indicated.
Except for the  bottom curve, the curves have  been vertically shifted
(three decades) in order to clarify the picture.}
\label{FIG9}
\end{figure}

\begin{figure}
\epsfig{file=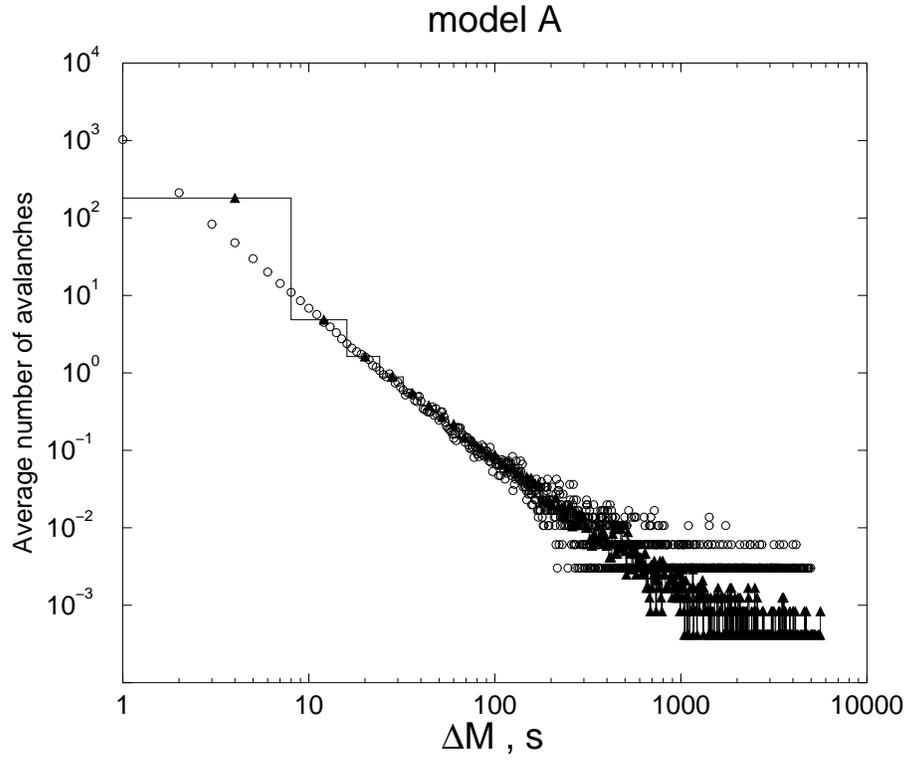, width=12cm}
\caption{Comparison of  the histograms $N(s)$  (circles) and $N(\Delta
M)$ (continuous line with black triangles indicating the centre of the
logarithmic bin)  corresponding to model  A with a system  size $L=20$
and  with $\theta_0=1.39$.   Data  correspond to  averages over  $300$
different realizations.}
\label{FIG10}
\end{figure}

\begin{figure}
\epsfig{file=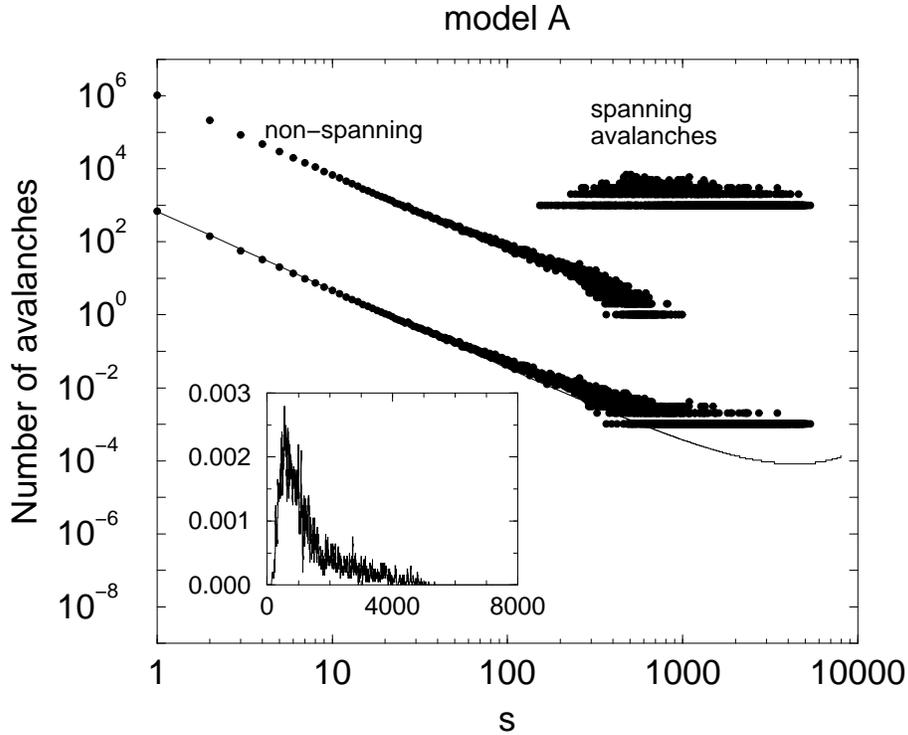, width=12cm}
\caption{Distribution  of  avalanches  for  model A  with  $L=20$  and
$\theta_0=1.39$.  Data  corresponds to averages  over $1000$ different
configurations of  disorder.  The bottom histogram  corresponds to the
analysis  of all avalanches.   The top  histogram (shifted  6 decades)
corresponds to the  spanning avalanches and the middle  one (shifted 3
decades)  to  the  non-spanning   avalanches.   The  inset  shows  the
histogram of spanning avalanches on a linear scale.  Data in the inset
has been smoothed in order to clarify the picture. }
\label{FIG11}
\end{figure}

\begin{figure}
\epsfig{file=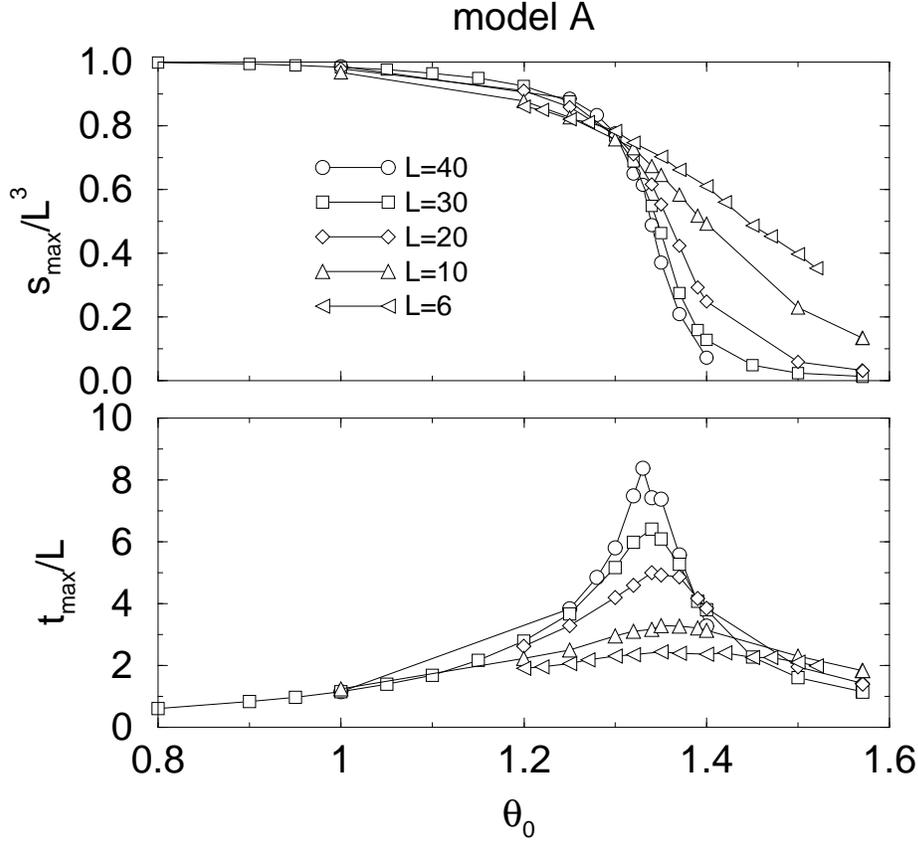, width=12cm}
\caption{Average size (a) and duration (b) of the largest avalanche in
the  full hysteresis  loop as  a function  of the  amount  of disorder
$\theta_0$ for model  A. Data correspond to different  system sizes as
indicated  by   the  legend  and  to  averages   over  many  different
configurations of disorder.}
\label{FIG12}
\end{figure}

\begin{figure}
\epsfig{file=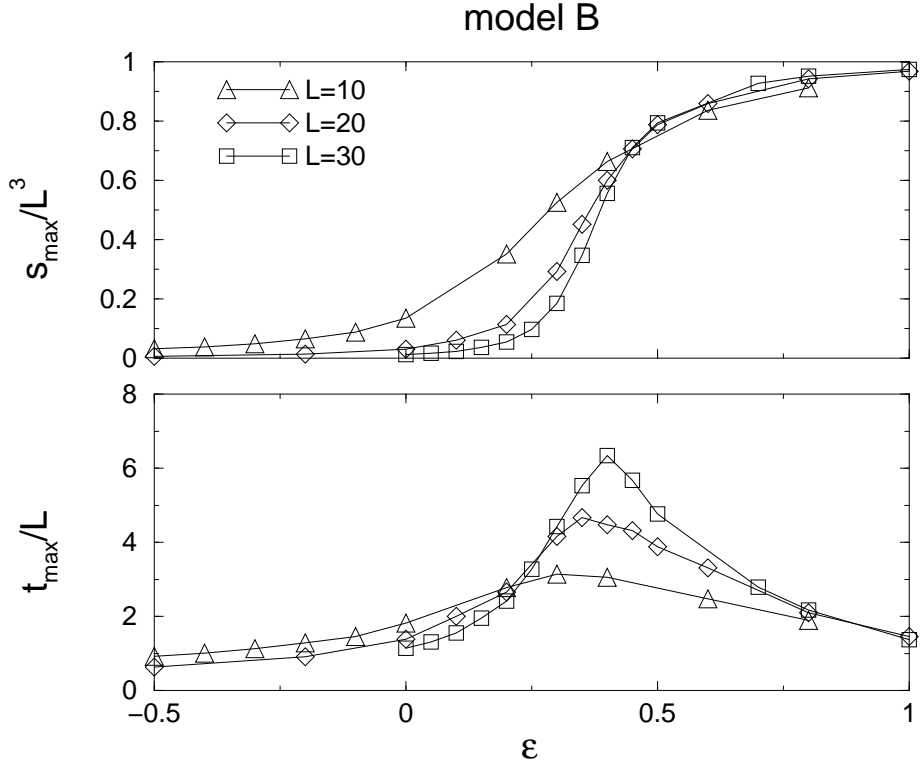, width=12cm}
\caption{Average size (a) and duration (b) of the largest avalanche in
the  full hysteresis  loop as  a function  of the  amount  of disorder
$\epsilon$ for model B.}
\label{FIG12B}
\end{figure}

\begin{figure}
\epsfig{file=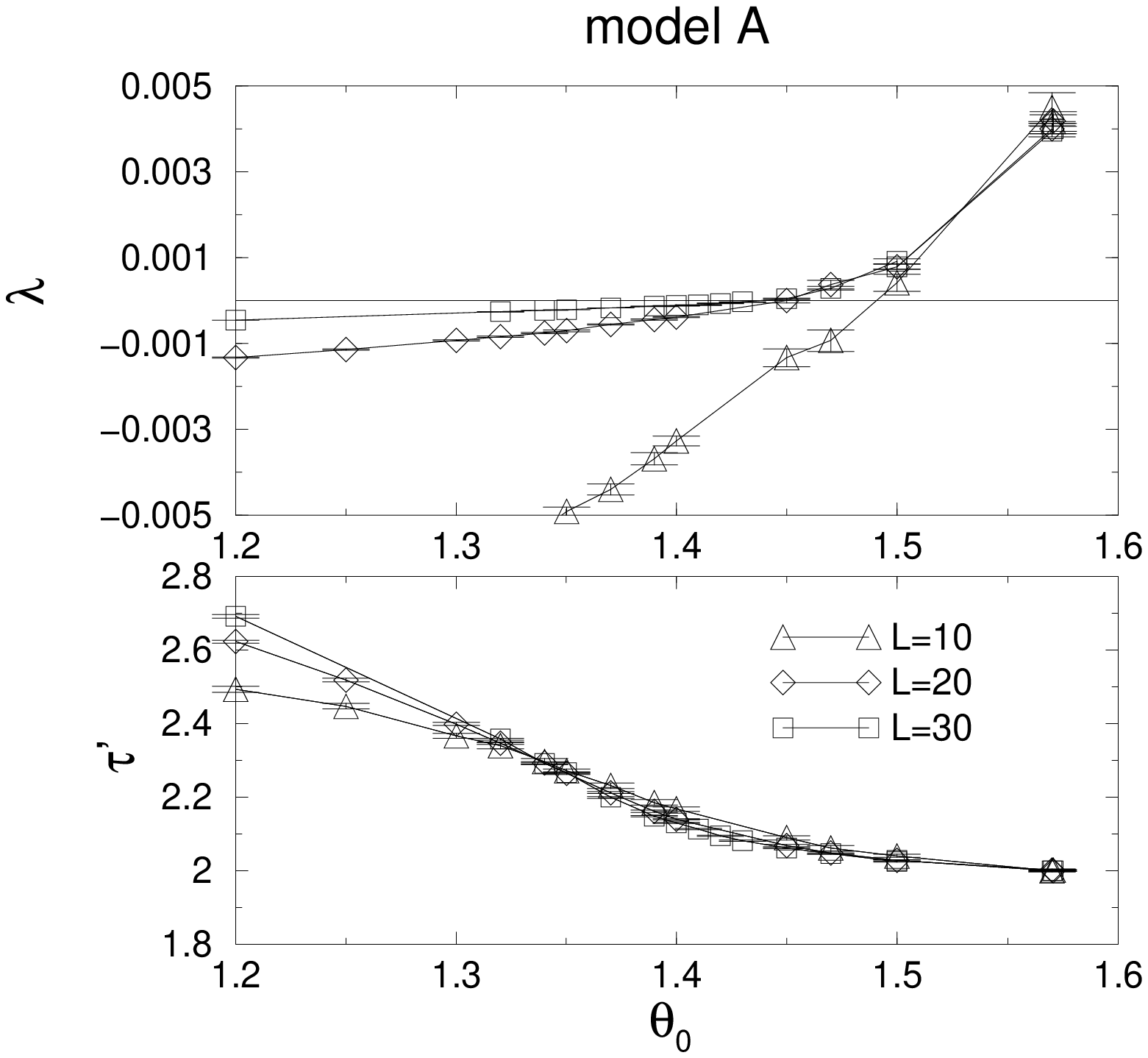, width=12cm}
\caption{Fitted parameters $\lambda$ and $\tau'$ to the avalanche size
distributions as  defined in equation  \ref{powexp} for model  A. Data
correspond to different system sizes as indicated by the legend and to
averages over many different configurations of disorder. }
\label{FIG13}
\end{figure}

\begin{figure}
\epsfig{file=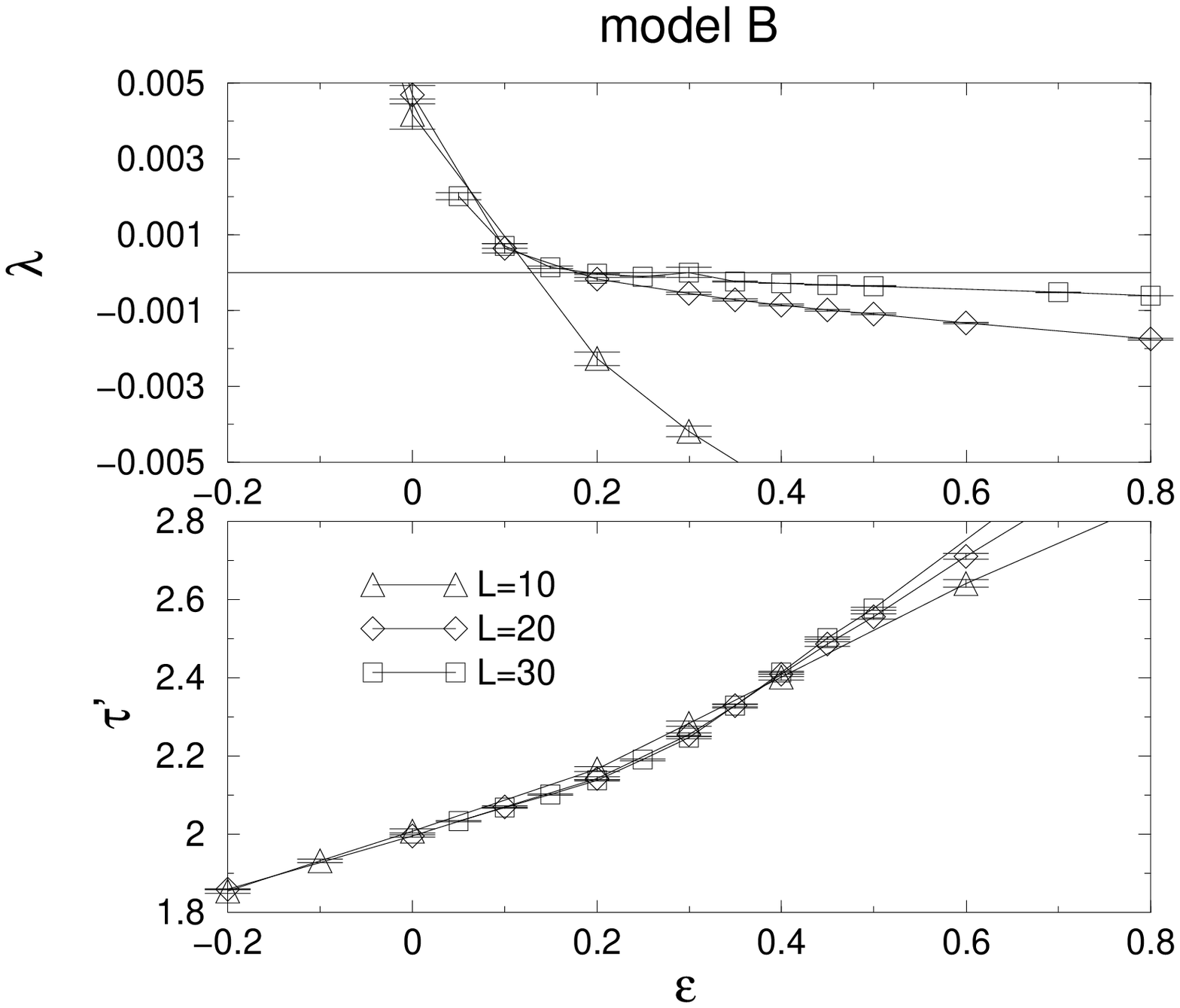, width=12cm}
\caption{Fitted parameters $\lambda$ and $\tau'$ to the avalanche size
distributions as  defined in equation  \ref{powexp} for model  B. Data
correspond to different system sizes as indicated by the legend and to
averages over many different configurations of disorder. }
\label{FIG13B}
\end{figure}

\begin{figure}
\epsfig{file=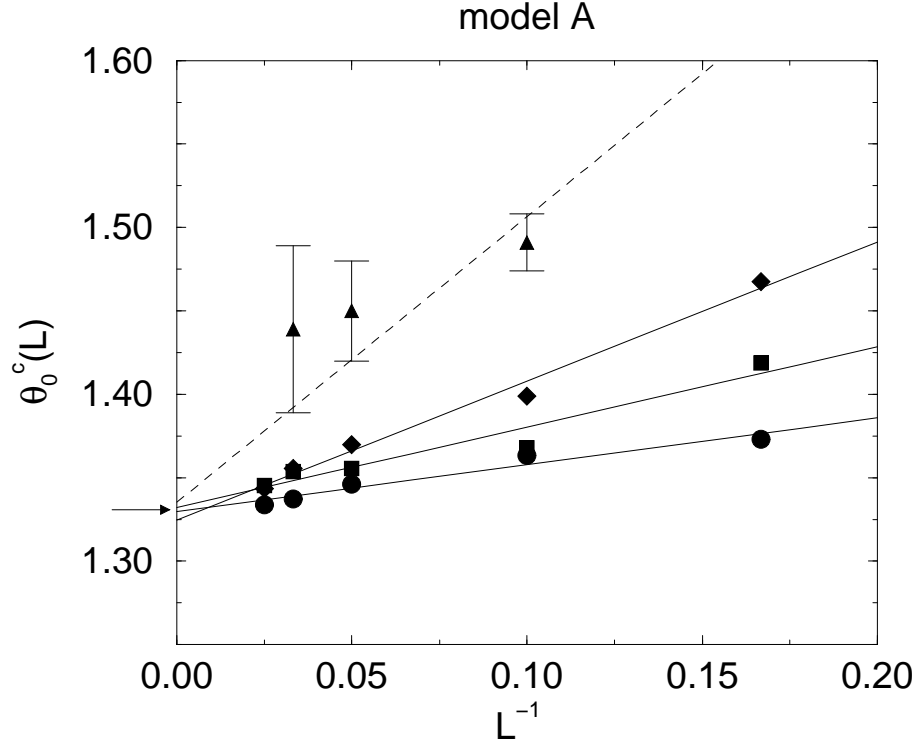, width=12cm}
\caption{Dependence  of different  estimations  of $\theta_0^c(L)$  on
$L^{-1}$ for model A.  Circles  correspond to the estimations from the
position  of  the  maximum  in $t_{max}(\theta_0)$,  diamonds  to  the
position of  the inflection  point in $s_{max}(\theta_0)$,  squares to
the  position   of  the  minimum   in  the  numerical   derivative  $d
s_{max}(\theta_0) / d \theta_0$,  and triangles to the disorder values
for   which  the  parameter   $\lambda$  vanishes.   Continuous  lines
correspond to linear fits used for the extrapolation to $L \rightarrow
\infty$. The dashed  line is a guide to the  eye.  The arrow indicates
the value $\theta_c=1.33$.}
\label{FIG14}
\end{figure}

\begin{figure}
\epsfig{file=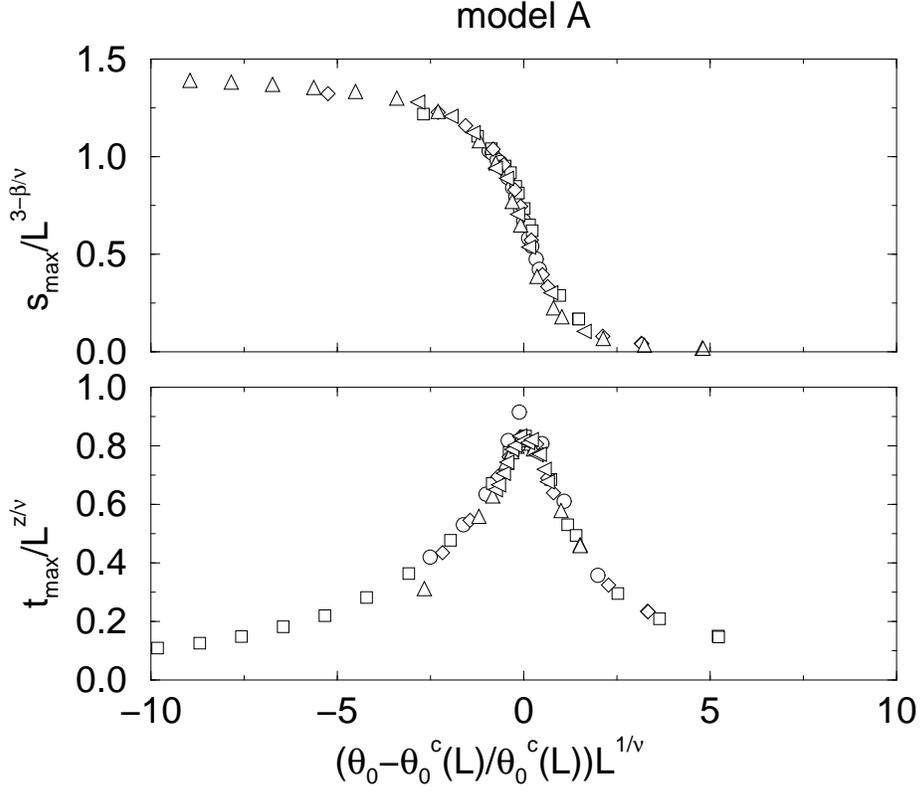, width=12cm}
\caption{Scaling of  the largest avalanche  size $s_{max}(\theta_0,L)$
and duration of the  longest avalanche $t_{max}(\theta_0,L)$ for model
A.  The values $\nu=1$, $\beta=0.1$ and $z=1.6$ have been used.}
\label{FIG15}
\end{figure}

\begin{figure}
\epsfig{file=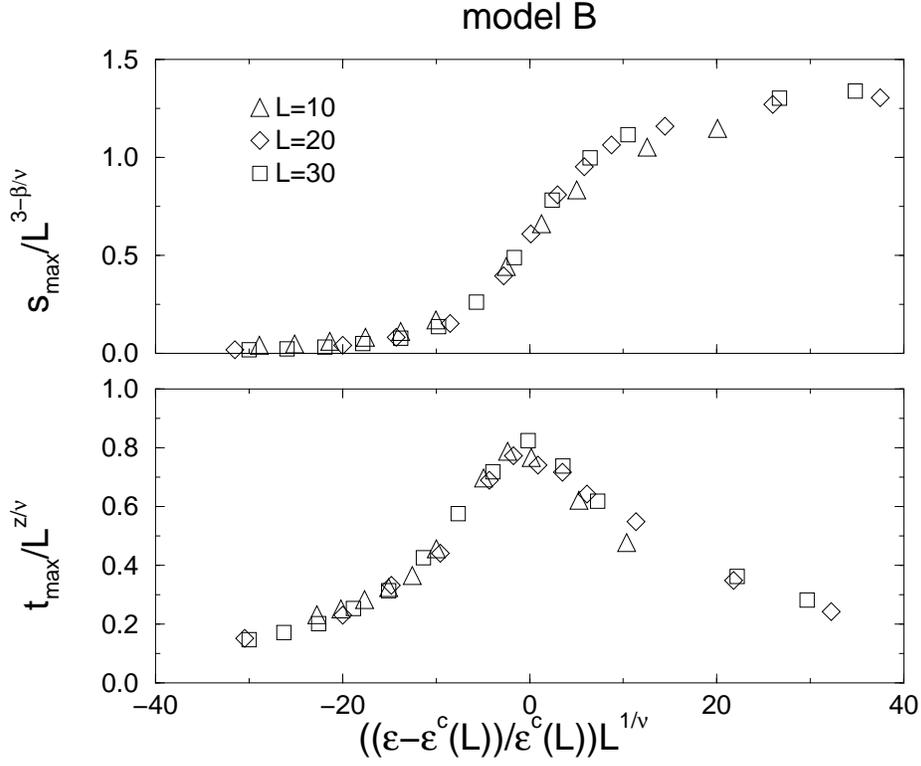, width=12cm}
\caption{Scaling of  the largest avalanche  size $s_{max}(\epsilon,L)$
and duration of the  longest avalanche $t_{max}(\epsilon,L)$ for model
B.  The values $\nu=1$, $\beta=0.1$ and $z=1.6$ have been used.}
\label{FIG15B}
\end{figure}

\begin{figure}
\epsfig{file=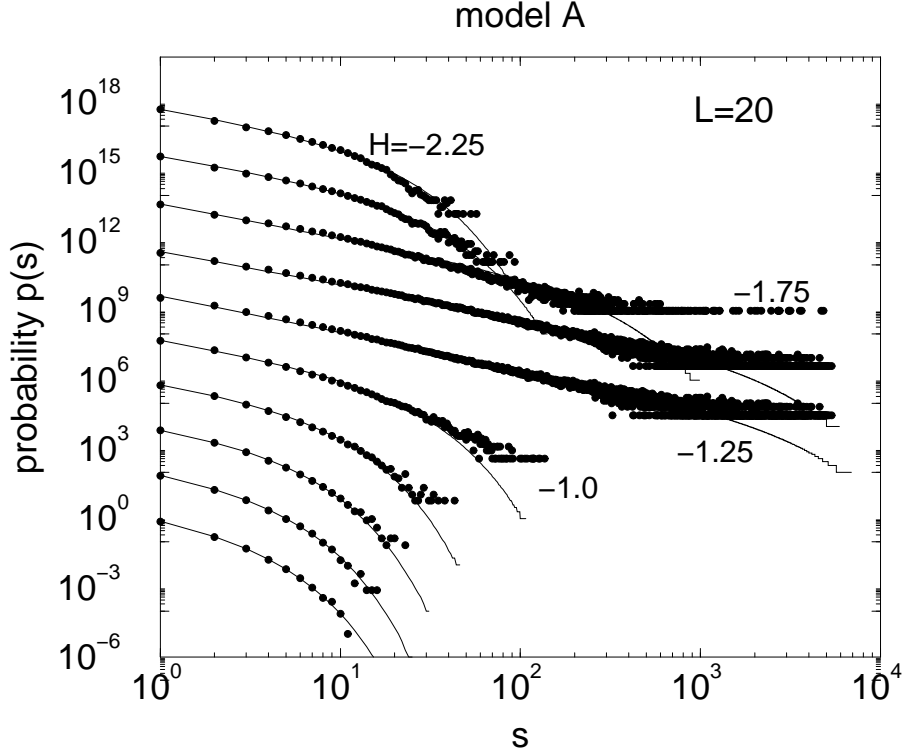, width=12cm}
\caption{Avalanche size distribution corresponding to different values
of the  applied external field  $H$ for model  A with size  $L=20$ and
$\theta_0=1.39$.   Histograms  have  been  performed by  counting  the
avalanches within  a window of $\Delta  H = 0.5$  centred on different
values of the  field.  From bottom to top such  fields vary from $0.0$
to  $-2.25$ with  steps  of $0.25$.   Moreover,  averages over  $1000$
realizations of  disorder have  been performed.  Histograms  have been
shifted two decades each in order to clarify the picture.}
\label{FIG16}
\end{figure}

\begin{figure}
\epsfig{file=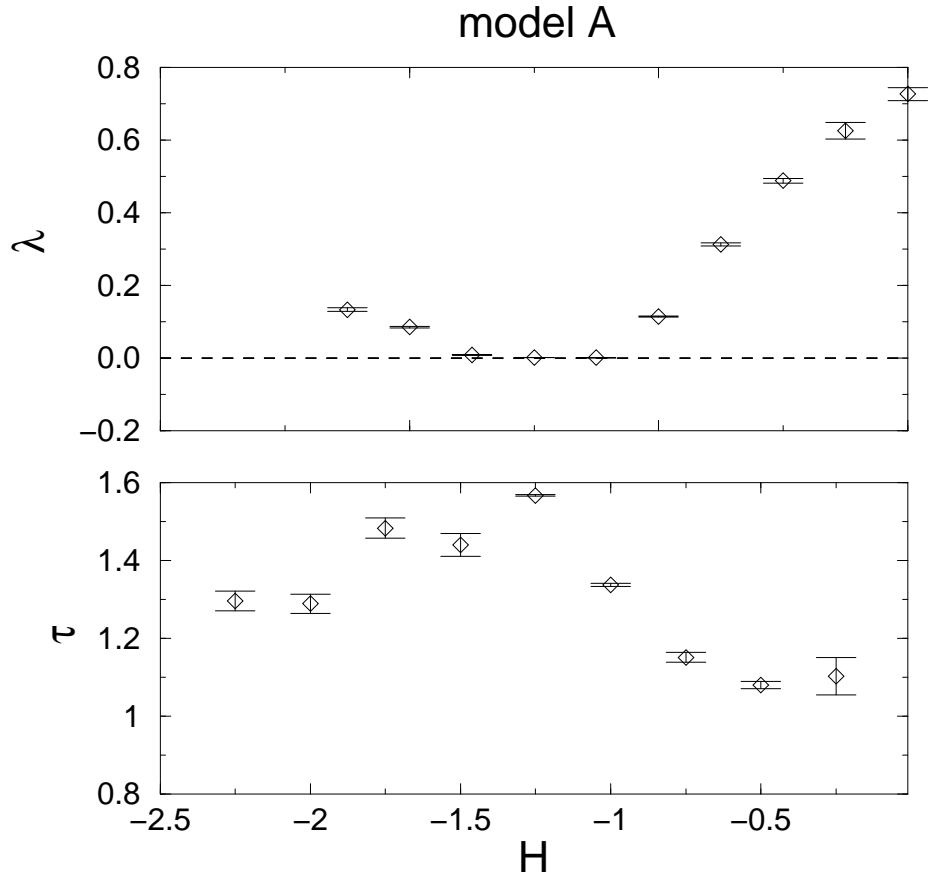, width=12cm}
\caption{Parameters $\lambda$ and $\tau$ as a function of the external
field  $H$   fitted  from   the  histograms  in   figure  \ref{FIG16},
corresponding to a system with size $L=20$ and $\theta_0=1.39$}
\label{FIG17}
\end{figure}

\begin{figure}
\epsfig{file=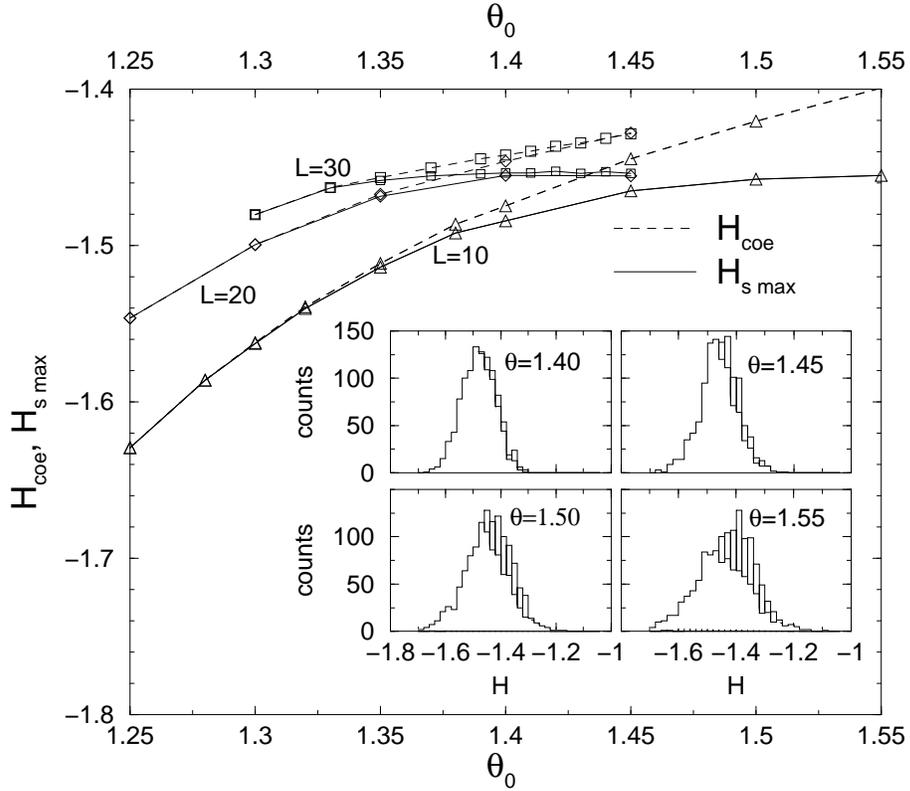, width=12cm}
\caption{Comparison of  the dependence of the  coercive field $\langle
H_{coe} \rangle $ and the field for which the largest avalanche occurs
$\langle H_{s_{max}} \rangle $ as a function of the amount of disorder
$\theta_0$ for  model A  with $L=10$, $20$  and $30$.  Inset  show the
actual distribution of the  two quantities ($\langle H_{coe} \rangle $
with  an empty  histogram and  $\langle H_{s_{max}}  \rangle $  with a
lined histogram) for  $L=10$ at four values of  the amount of disorder
as indicated.}
\label{FIG18}
\end{figure}

The  morphological  properties  of  the hysteresis  loops  depend,  as
expected,  on the  specific  characteristics of  the  disorder. It  is
interesting  to  compare the  results  given  in  Fig. \ref{FIG5}  for
coercivity and dissipated  energy with available experimental results.
Experiments carried out on  ribbons of high magnetostrictive amorphous
alloys    under   stress   \cite{Appino99,Durin99}    are   especially
interesting.   They  reveal that  the  applied  stress favours  global
(long-range) uniaxial anisotropy which manifests itself by a change of
the magnetic  domain pattern \cite{Livingston85}.   Consequently, this
leads  to a  change of  the  shape of  the hysteresis  loops.  At  low
external  stress, a  complicated pattern  constituted by  maze domains
results  from the  effect of  quenched-in stresses.   As  the external
stress  is  increased,  a  simpler  domain pattern  appears  with  few
parallel domains in the  direction of the external stress.  Therefore,
it seems  reasonable to  assume that  the effect of  the stress  is to
reduce the randomness of the  local anisotropy axes, or in other words
to reduce  disorder.  The main experimental  result \cite{Durin99}, is
that with  increasing external stress, $H_{coe}$  initially exhibits a
fast decrease down  to a certain minimum value,  followed by a roughly
linear increase at high stresses.  Actually, this effect is reproduced
by our  model B as can be  seen in Fig.  (\ref{FIG5}).   This could be
explained  by   taking  the  competition  phenomena   arising  in  the
$\epsilon<0$ region  for model  B mentioned in  section \ref{modeling}
into  account.  More quantitative  comparisons, nevertheless,  are not
possible since  the actual anisotropy axis distribution  in ribbons is
difficult   to  compare  with   our  3d   system.   As   regards  $W$,
experimentally it is  found that it shows a  behaviour similar to that
of $H_{coe}$  \cite{Appino99}, that  is, it exhibits  a minimum  for a
certain value of the applied stress.  This is not reproduced either by
our models  A or B, which  show simply a monotonous  increase when the
system becomes more and more  ordered.  This disagreement could be due
to the  fact that  in the  models, $M_{sat}$ depends  on the  degree of
disorder,  as  a  consequence  of the  strong  anisotropy  assumption.
Experimentally this is  not the case and $M_{sat}$  is almost constant
for a  given sample composition and  thus one expects  that $W \propto
H_{coe}$.

As regards the critical point  our results are totally compatible with
the  universality   that  has  been  proposed   for  similar  athermal
models. Table \ref{TABLE1} compares the values obtained in the present
work for models A and B  with those reported in the literature for the
3d  Random Field  Ising model  ,  3d Random  Bond Ising  model and  3d
Site-Diluted  Ising   model.   The  agreement   is  very  satisfactory
confirming  universality  \cite{Dahmen96,Vives95}. Table  \ref{TABLE1}
also includes the values  of the exponents corresponding to mean-field
calculations.   Clearly, when  considering  the full  set  of all  the
critical exponents,  one concludes that the mean-field  model does not
belong to the  same universality class.  This is  not surprising since
the mean-field approximation assumes long-range interactions while the
other models are strictly  short range. The mean-field exponent values
are  expected to be  found in  systems including  dipolar interactions
\cite{Kuntz99}.  Nevertheless it should be remarked that the exponents
$\tau$ and $\tau'$ seem to  have, within the errors, comparable values
for  the two  universality classes.   Therefore, the  analysis  of the
models suggests that the  statistical distribution of avalanches shows
very close critical exponents, irrespective of the inclusion or not of
the dipolar forces.

It is, perhaps, more interesting to compare such theoretical exponents
with  those  found  experimentally.   The  direct  comparison  of  the
numerical  values   should  always   be  taken  carefully   since,  in
experiments,  avalanche   sizes  are  determined   in  different  ways
depending   on  the  specific   measurement  technique   used.   Table
\ref{TABLE2} summarizes  the most significant values  of the exponents
reported from  the experimental study of Barkhausen  noise in magnetic
systems   \cite{Lieneweg72,Spasojevic96,Cote91,Urbach95,Durin00}.   We
have separated  the $\tau$ exponents corresponding  to measurements of
noise around  a certain value of  the external field  from the $\tau'$
exponents corresponding to the  analysis of the signal sequence during
the full hysteresis loop (or half loop).  In both cases, the numerical
procedure for obtaining such experimental exponents is similar to that
followed for  the analysis  of the model  simulations; it is  based on
fitting  an  expression  like  Eq.   (\ref{powexp})  to  the  recorded
histograms of avalanche sizes.

A  first   remark  is  that  the  overall   situation  concerning  the
possibility for universality  in experiments is not as  clear as it is
for the  theoretical models.   In our opinion  the main problem  is to
determine whether  the analyzed data corresponds to  a critical system
or  not.   A  second remark  is  that  the  values reported  in  Table
\ref{TABLE2} seem to show a  certain dependence on heat treatments and
other  effects influencing  the  degree of  quenched  disorder in  the
system.    For   instance,   in   references   \cite{Lieneweg72}   and
\cite{Cote91} it is found  that the $\tau'$ exponent increases towards
a  value close to  $2.0$ when  the degree  of order  in the  sample is
increased by annealing and/or magnetic field cooling.  Furthermore the
distribution of avalanche sizes in Fe-Co-B alloys (characterized by an
exponent  $\tau=1.27$  \cite{Durin00}  )  were found  to  change  from
subcritical   towards  critical   (the  cutoff,   equivalent   to  our
$\lambda^{-1}$,  increases)   when  the  applied   tensile  stress  is
increased  \cite{Durin99}.   In  agreement  with  these  results,  the
$\tau'$ exponents  in our simulations  show a clear increase  when the
degree of disorder  is decreased as can be  seen in Figs.  \ref{FIG13}
and \ref{FIG13B}.   Moreover, our  results also suggest  that provided
that  the measurements are  performed in  the subcritical  region, the
estimated value  of $\tau'$ will  remain close to the  critical value.
This could explain why the values of $\tau'$ in Table \ref{TABLE2} are
quite similar to  those found for the models.   Actually, the possible
existence of a  large critical region has also  been suggested for the
RFIM \cite{Dahmen96} and for the Site Diluted Random Field Ising model
(RFIM with vacancies) \cite{Tadic96}.  For this last model it has even
been proposed  that true  criticality extends over  a broad  region of
parameters controlling disorder.

A third remark concerns certain  procedures used for the estimation of
$\tau$ and  $\tau'$.  For instance, for the  determination of $\tau'$,
in  some cases  saturation is  not reached  in the  studied hysteresis
loops.  This means  that the distributions correspond, in  fact, to an
internal loop. True saturation requires  very high fields which can be
not experimentally accessible.   At present, it is not  clear what the
consequences  on  the measured  $\tau'$  exponent  will  be.  For  the
determination  of $\tau$,  the  experiments are  carried  out with  an
external field constrained around the coercive field.  Our simulations
suggest that  this may  introduce a bias  in the  estimated exponents.
First, if the  amount of disorder is greater  than the critical amount
of disorder, the field at  which the largest avalanche takes place and
the   coercive  field   do   not   coincide,  as   can   be  seen   in
Fig. \ref{FIG18}.   Moreover, even  in the case  that the  disorder is
close to the critical value, a deviation in the tuning of the external
field would lead to lower  values of $\tau$ compared to those expected
at $H=H^c$, as  can be seen in Fig. \ref{FIG17}.   This may provide an
explanation for  some of  the low values  of $\tau$  reported recently
\cite{Durin00}.

\section{Summary and conclusion}

In this  paper we have studied  rate-independent hysteresis properties
of  a  Random Anisotropy  reticular  model.   We  have considered  the
infinite uniaxial  anisotropy limit and we have  neglected any effects
of dipolar interactions. In this limit the model reduces to the Random
Anisotropy  Ising Model  which can  be viewed  as a  combination  of a
Random  Bond Ising  Model  with random  couplings  (g-factors) to  the
external  magnetic  field.  This  model  seems  rather appropriate  to
realistically   describe   amorphous   and  polycrystalline   magnetic
materials.  Disorder  is introduced in the system  through the spatial
random distribution of  anisotropy axis.  Two different distributions,
in which disorder is controlled  by a single parameter ($\theta_0$ and
$\epsilon$), have been  considered. Extensive numerical simulations of
the model  have been performed  by means of a  deterministic algorithm
consisting   in  of  synchronous   local  relaxation   dynamics.   The
morphological properties  of the hysteresis loops  have been obtained.
They depend on  the specific distribution of disorder  but do not show
any  singular  behaviour when  $\theta_0$  or  $\epsilon$ are  varied.
Qualitative agreement  with some available experimental  data has been
found.   We  expect  that  by  choosing  a  suitable  phenomenological
distribution of disorder such morphological properties could be better
reproduced.

Besides, we have focused on  the analysis of the Barkhausen avalanches
generated   during   the   metastable  evolution.    The   statistical
distribution  of such  avalanches  shows a  critical  behaviour for  a
certain  amount of  disorder ($\theta_0^c  \simeq 1.33  \pm  0.03$ and
$\epsilon^c  \sim  0.2$).  From  a  finite-size  scaling  analysis  of
different simulated properties we have obtained the critical exponents
characterizing the disorder-induced critical point. The most important
conclusion is  that the present  model falls in the  same universality
class of the athermal 3d RFIM.

We have  also analyzed the different available  experimental values of
such critical exponents  characterizing the distribution of Barkhausen
signals. Data is  scarce and refer to different  exponents ($\tau$ and
$\tau'$).   Although  there  are  some discrepancies,  the  comparison
indicates  that  the  experimental  systems  may fall  into  the  same
universality class.  However, results  suggest that it is necessary to
tune  the  disorder  in  the systems  with  adequate  thermomechanical
treatments  so  that  the   system  behaves  critically.   Although  a
systematic  control  of  the  amount  of  disorder  is  experimentally
difficult, the analysis of our model indicates the best conditions for
such  measurements  and  data  analysis; (i)  disregarding  additional
experimental  problems  \cite{Bertotti81},  it  is  more  reliable  to
measure $\tau'$  (full hysteresis loop analysis) instead  of $\tau$ in
order to avoid  problems related to the determination  of the critical
field $H_c$; (ii) although the samples exhibit an exponentially damped
power-law distribution  (subcritical), the $\tau'$  exponents obtained
by fitting Eq.  \ref{powexp}  render good estimations of the exponents
at criticality.  Therefore, measurements in the subcritical region are
preferable to measurements in the supercritical region.

\acknowledgements

This work has received financial support from the CICyT (Spain), project
MAT98-0315 and from the CIRIT (Catalonia), project 1998SGR48.

\newpage

\begin{table}
\begin{tabular}{l|l|l|l|l|l|}
Model &  $\beta$ & $\tau'$  & $\tau$ &  $z$ & $\nu$ \\  \hline 3d-RAIM
(model A) &  $0.1\pm 0.1$ & $2.06 \pm 0.05$ &  &$1.6 \pm 0.1$ &$1.0\pm
0.1$\\ \hline 3d-RAIM (model B) & $0.1^*$ & $2.10 \pm 0.05$ & &$1.6^*$
&$1.0^*$  \\  \hline  3d-RFIM  \cite{Perkovic99}  &  $0.035\pm  0.028$
&$2.03\pm  0.03$  &$1.6\pm  0.06$   &  &  $1.41  \pm  0.17$\\  3d-RBIM
\cite{Vives95}  & $0.0\pm  0.1$  & $2.0\pm  0.2$  & &  $1.6\pm 0.1$  &
$1.06\pm 0.1$ \\ 3d-SDIM \cite{Vives00} &  &$1.9\pm 0.2$ & & & \\ Mean
Field \cite{Perkovic99} &$1/2$ & $2$ &$3/2$ & &$1/2$ \\
\end{tabular}
\caption{Critical  exponents from numerical  simulations in  this work
and in  the literature. The values  with an asterisk  ($^*$) have been
obtained from  model A data and  were checked for scaling  the data of
model B.}
\label{TABLE1}
\end{table}

\begin{table}
\begin{tabular}{|c|c|c|c|c|c|}
Material &  Heat treatments  & $\tau'$  & $\tau$ &  Observ. &  Ref. \\
\hline 
 $81\%$Ni-Fe & $1h$ at  $240^o C$ & $1.73$ & & & \cite{Lieneweg72} \\ 
\cline{2-4} 
 & $1h$ at $460^o C$ & $2.1$ & & & \\ 
\hline
\begin{minipage}{3.5cm}
\vspace{2mm}
\begin{center}
VITROVAC 6025-X  \\  (metal-glass)
\end{center}
\vspace{2mm}
\end{minipage}
& & & $1.77$ & \parbox{4cm}{small internal loops } &\cite{Spasojevic96}\\
\hline 
Metglass 2605S-2 & as cast & $1.85$ & &
\begin{minipage}[c]{4cm}
\vspace{2mm}
\begin{center}
$\tau'$  calculated from \\ scaling relations \\ and other measured \\ exponents 
\vspace{2mm}
\end{center}
\end{minipage}
&
\cite{Cote91}\\ 
\cline{2-4} 
& 
\begin{minipage}{3.6cm}
\vspace{2mm}
annealed at $400^oC$ \\ and field cooled \\ ($25^o C/min$, $120 Oe$)
\vspace{2mm}
\end{minipage} 
& 2.0 & &  & \\
\hline 
\begin{minipage}{3.8cm}
\vspace{2mm}
\begin{center}
Perminvar \\
30$\%$Fe 45$\%$Ni 25$\%$Co
\vspace{2mm}
\end{center}
\end{minipage} &
\begin{minipage}{3.6cm} \vspace{2mm} Annealed $1 h $, $1000^o C$ \\ $24 h$ $450^o C$ \vspace{2mm} \end{minipage} 
& &$1.33$ & & \cite{Urbach95} \\
\hline
Fe-Si 7.8 wt\%  & Annealed $950^o C$ & & $1.5 \pm 0.05$  & Polycrystalline & \cite{Durin00} \\
\cline{1-2}
Fe-Si 6.5 wt\%  & Annealed $1200^o C$ & & & & \\
\cline{2-2}
 & Annealed $1050^o C$ & & & & \\
\cline{1-5}
\begin{minipage}{4cm}
\vspace{2mm}
\begin{center}
Fe$_{21}$ Co$_{64}$ B$_{15}$ \\
Fe$_{64}$ Co$_{21}$ B$_{15}$
\end{center}
\vspace{2mm}
\end{minipage}

& as cast & & $1.27 \pm 0.03$ & 
\begin{minipage}{4cm}
\vspace{2mm}
\begin{center}
Amorphous \\
under stress
\end{center}
\vspace{2mm}
\end{minipage}
& \\
\end{tabular}

\caption{Experimental  values  of the  critical  exponents $\tau$  and
$\tau'$ }
\label{TABLE2}
\end{table}

\newpage

\newpage

\newpage

\newpage

\newpage

\newpage

\newpage

\newpage

\newpage

\newpage

\newpage

\newpage

\newpage

\newpage

\newpage

\newpage

\newpage

\newpage

\newpage

\newpage

\newpage

\end{document}